\documentclass[a4paper,usenatbib]{mnras}
\usepackage{aas_macros}
\usepackage{amsfonts}
\usepackage{amsmath}
\usepackage{amssymb}
\usepackage{graphicx}
\usepackage[dvipsnames]{xcolor}
\hypersetup{colorlinks=true,allcolors=teal}
\usepackage{microtype}

\newcommand{\NSIM}{\ensuremath{15}}

\newcommand{\p}{\ensuremath{\partial}}

\newcommand{\Msun}{\ensuremath{M_{\odot}}}
\newcommand{\Mh}{\ensuremath{h^{-1}M_{\odot}}}

\newcommand{\Mpch}{\ensuremath{h^{-1}{\rm Mpc}}}

\newcommand{\avg}[1]{\ensuremath{\left\langle \,#1\, \right\rangle}}
\newcommand{\e}[1]{\ensuremath{{\rm e}^{#1}}}

\newcommand{\der}{\ensuremath{{\rm d}}}

\newcommand{\dL}{\ensuremath{\delta_{\rm L}}}
\newcommand{\dHL}{\ensuremath{\delta_{\rm h}^{\rm L}}}
\newcommand{\dH}{\ensuremath{\delta_{\rm h}}}

\newcommand{\eqn}[1]{equation~\eqref{#1}}
\newcommand{\eqns}[1]{equations~\eqref{#1}}

\newcommand{\ph}[1]{\phantom{#1}}

\newcommand{\be}{\begin{equation}}
\newcommand{\ee}{\end{equation}}
\newcommand{\Cal}[1]{\ensuremath{\mathcal{#1}}}

\begin{document}
\label{firstpage}
\pagerange{\pageref{firstpage}--\pageref{lastpage}}

\title[Separate Universe assembly bias]{Halo assembly bias from Separate Universe simulations}

\author[Paranjape \& Padmanabhan]{
Aseem Paranjape$^{1}$\thanks{E-mail: aseem@iucaa.in}
and Nikhil Padmanabhan$^{2}$\thanks{E-mail: nikhil.padmanabhan@yale.edu}
\\
$^{1}$Inter-University Centre for Astronomy and Astrophysics, Ganeshkhind, Post Bag 4, Pune 411007, India\\
$^{2}$Department of Physics, Yale University, New Haven, CT 06520\\
}

\date{draft}

\maketitle 

\begin{abstract}
\noindent
We present a calibration of halo assembly bias using the Separate Universe technique. 
Specifically, we measure the response of halo abundances at fixed mass and concentration to the presence of an infinite-wavelength initial perturbation. 
We develop an analytical framework for describing the concentration dependence of this peak-background split halo bias -- a measure of assembly bias -- relying on the near-Lognormal distribution of halo concentration at fixed halo mass. 
The combination of this analytical framework and the Separate Universe technique allows us to achieve very high precision in the calibration of the linear assembly bias $b_1$, and qualitatively reproduces known trends such as the monotonic decrease (increase) of $b_1$ with halo concentration at large (small) masses. 
The same framework extends to the concentration dependence of higher order bias parameters $b_n$, and we present 
the first calibration of assembly bias in $b_2$. 
Our calibrations are directly applicable in analytical Halo Model calculations that seek to robustly detect \emph{galaxy} assembly bias in observational samples. 
We detect a non-universality in the $b_1 - b_2$ relation arising from assembly bias, and suggest that simultaneous measurements of these bias parameters could be used to both detect the signature of assembly bias as well as mitigate its effects in cosmological analyses. 
\end{abstract}

\begin{keywords}
cosmology: theory, dark matter, large-scale structure of the Universe -- methods: numerical, analytical
\end{keywords}

\section{Introduction}
\label{sec:intro}
\noindent
The correlation between dark matter halo properties and halo environment is of great interest in understanding the formation of large scale structure in the Universe \citep[for a recent review, see][]{djs17}. Early work using analytical models predicted that halo mass is the primary variable that correlates with environment; the clustering strength of massive haloes relative to dark matter is a strong function of halo mass \citep{Kaiser84,bbks86,mw96,st99}. $N$-body simulations of cold dark matter (CDM) have also shown that halo clustering additionally depends on the \emph{assembly history} of haloes at fixed mass \citep{st04,gsw05}. This `halo assembly bias' manifests as a dependence of the large scale clustering of haloes on a variety of halo properties that correlate with assembly history, including halo concentration, shape, spin, velocity anisotropy, etc. \citep{wechsler+06,jsm07,fm10,fw10}.

On the observational front, the relevance of assembly bias for \emph{galaxy} properties has been more difficult to establish. Semi-analytic models of galaxy evolution routinely assume a tight correlation between the star formation history of galaxies and the mass accretion rates of CDM haloes, thus predicting a galaxy assembly bias when galaxies are split, for example, by colour or specific star formation rate at fixed stellar mass \citep[see][for a review]{sd15}. Statistical modelling of galaxy abundances and clustering -- e.g., the Halo Occupation Distribution approach \citep[HOD;][]{Seljak2000,sshj01,bw02} -- on the other hand, has traditionally assumed that galaxy properties are fully determined by halo mass alone, and hence that there is \emph{no} galaxy assembly bias. It has been challenging to distinguish between these models in observational samples, primarily because the low-mass regime where one expects a strong signature involves faint objects which are difficult to observe and classify robustly \citep{lin+16}. While there have been recent claims of assembly bias for massive clusters \citep{miyatake+16}, the observed signal is possibly consistent with systematic effects in the cluster selection \citep{zu+17}. Observational constraints in the low mass regime are still inconclusive \citep{twcm16}.

There is, therefore, considerable interest in developing analytical and statistical tools that incorporate halo assembly bias and allow for a potential effect on galaxy properties. Analytical models of CDM dynamics based on peaks theory \citep{dwbs08}, ellipsoidal dynamics \citep{desjacques08} and modern versions of the excursion set approach \citep{ms12,cs13} predict that, at fixed mass, haloes that assembled \emph{early} -- and, consequently, are highly concentrated in their current density profile -- should cluster \emph{less strongly} than late forming, low concentration haloes of the same mass. While the measurements of massive haloes in simulations are indeed qualitatively consistent with this predicted trend, the situation at the low mass end is quite different. For haloes substantially smaller than a characteristic mass scale, the measured assembly bias trend \emph{inverts}. Early forming low mass haloes are \emph{more} clustered than their late forming counterparts. This inversion is very likely due to the tidal influences of massive objects on low mass haloes in filaments \citep{hahn+09,bprg16}. As yet, however, there is no robust analytical understanding of this inversion of the assembly bias trend.

In this backdrop, there is clearly a need for accurate calibrations of the halo assembly bias signature which might then be used in Halo Model analyses of galaxy properties. Despite the large number of studies of halo assembly bias, there have been remarkably few attempts to accurately calibrate the effect and eventually incorporate it in an extended HOD framework \citep{wechsler+06}. In this paper, we address the issue of calibration using the Separate Universe technique \citep{tb96,cole97,bssz11,wsck15a,lht16}, which has been recently demonstrated to be an excellent noise-reduction method for measuring large scale halo bias, giving an exact realisation of the peak-background split \citep{lwbs16}. As we will show below, the Separate Universe technique is also ideally adapted to be combined with an analytical formalism for describing the dependence of halo bias on halo properties at a given epoch (we will focus on halo concentration). As an application, we will describe how the combined properties of assembly bias at linear and quadratic order can be of cosmological interest.

The paper is organised as follows. In section~\ref{sec:sims}, we summarize the Separate Universe technique and describe our simulations and analysis methods. In section~\ref{sec:biasSU}, we first develop an analytical formalism that isolates the concentration dependence of halo bias $b_n$ at generic order $n \geq 1$ from measurements in the Separate Universe simulations, and present results for our calibration of assembly bias for $n=1,2$. We discuss our results in the context of the current understanding of assembly bias in section~\ref{sec:discuss}, and finally summarize in section~\ref{sec:conclude}. The Appendices give technical details of some of the results used in the text.

\section{Simulations}
\label{sec:sims}
\noindent
In this section we describe our Separate Universe simulations and discuss our techniques for halo finding and measurements of halo concentrations.

\subsection{The Separate Universe technique}
\label{subsec:SUtech}
\noindent
The Separate Universe technique relies on the exact mapping between the cosmology described by an infinite-wavelength initial perturbation \dL\ on an existing, fiducial Friedmann-Lema\^itre-Robertson-Walker (FLRW) spacetime, and an FLRW cosmology with the same physical matter density, but a different spatial curvature and Hubble constant that are related to \dL\ in a specific way. The existence of this mapping means that, by running $N$-body simulations with carefully chosen (non-flat) background cosmologies, one can measure the response of evolved cosmological quantities (such as the halo mass function) to \dL. Specifically, if \dL\ denotes the initial perturbation extrapolated to present day using linear theory in the fiducial cosmology, then this response, at order $\dL^n$, is precisely what is meant by the $n^{\rm th}$ order peak-background split Lagrangian bias parameter $b_n^{\rm L}$. In other words, once the mapping between \dL\ and FLRW cosmology is in place, the Separate Universe technique provides us with a formally exact (and very accurate) measurement technique for the scale-independent, peak-background split halo bias.

These ideas have been recently implemented by \citet{lwbs16} for obtaining calibrations of halo bias at order $n\leq3$ with very high precision. In the present work, we will follow \citet{wsck15a} to set up and run our Separate Universe simulations, and \citet[][hereafter, L16]{lwbs16} to analyse them and measure the linear and quadratic bias coefficients. We refer the reader to \citet{wsck15a} for details of the 
$\dL\to$ FLRW mapping discussed above.
Our fiducial cosmology is a flat $\Lambda$CDM model with total matter density parameter $\Omega_{\rm m}=0.276$, baryonic matter density $\Omega_{\rm b}=0.045$, Hubble constant $H_0=100h\,{\rm kms}^{-1}{\rm Mpc}^{-1}$ with $h=0.7$, primordial scalar spectral index $n_{\rm s}=0.961$ and r.m.s. linear fluctuations in spheres of radius $8\Mpch$, $\sigma_8=0.811$. We will quote masses in \Msun\ and lengths in Mpc, with the understanding that we have appropriately converted outputs of codes that work with \Mh\ and \Mpch\ by default.

\begin{figure*}
\centering
\includegraphics[width=0.9\textwidth]{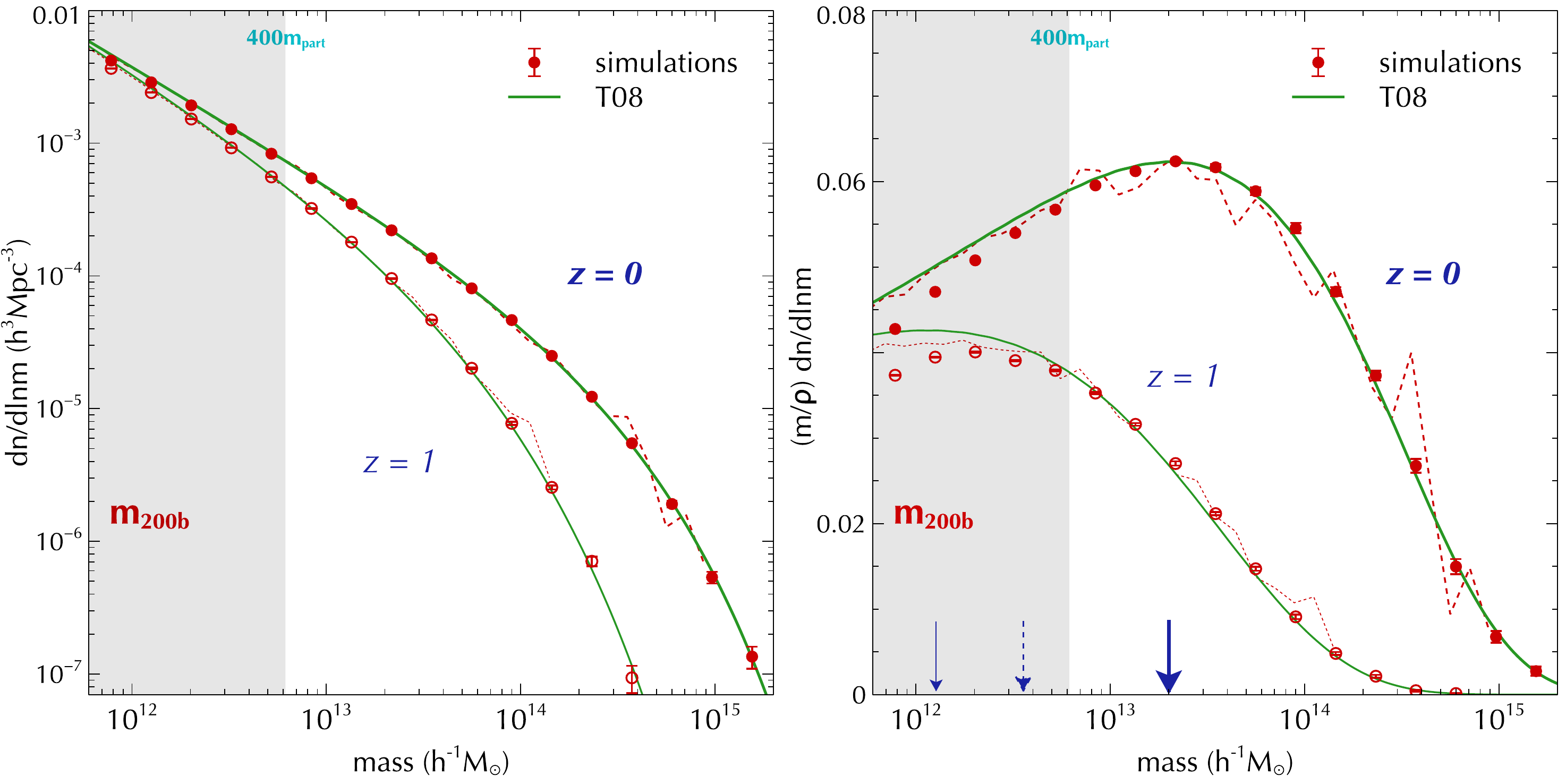}
\caption{Mass functions \emph{(left panel)} and mass fractions \emph{(right panel)} at $z=0$ (filled symbols) and $z=1$ (empty symbols) averaged over \NSIM\ realisations of the $512^3$-particle simulation with the fiducial cosmology. The dashed (dotted) lines show the corresponding $z=0$ ($z=1$) measurements in the $1024^3$-particle simulation. The solid lines show corresponding results using the fitting function from \citet[][T08]{Tinker08}. The thick (thin) solid arrow in the right panel marks the mass scale $m_\ast(z)$ at which the $z=0$ ($z=1$) T08 mass fraction peaks, while the dashed arrow shows the value $m_{\ast,{\rm std}}(z=0)$ obtained from the standard definition $\sigma(m_{\ast,{\rm std}})=\delta_{\rm sc}$. The shaded region indicates the masses we discard in our analysis of the Separate Universe simulations (see text for details).}
\label{fig:mf}
\end{figure*}

Our simulations all have the same comoving size, corresponding to $L_{\rm box}=(300/0.7)$Mpc, and contain $512^3$ particles each. Consequently, since the physical matter density parameter $\Omega_{\rm m}h^2$ remains constant across the simulations, the particle mass in each simulation is $m_{\rm part}=2.2\times10^{10}\Msun$. The simulations were performed using \textsc{gadget-2} \citep{springel:2005}\footnote{http://www.mpa-garching.mpg.de/gadget/} on the Perseus cluster at IUCAA\footnote{http://hpc.iucaa.in}, using a $1024^3$ PM grid and force resolution $\epsilon=27.9$kpc in comoving units, corresponding to $1/30$ of the mean inter-particle spacing. 

For exploring the space of Separate Universes related to our fiducial cosmology, we use a grid of values for the linearly extrapolated density contrast \dL\ given by $\dL\in$ \{$\pm$0.7, $\pm$0.5, $\pm$0.4, $\pm$0.3, $\pm$0.2, $\pm$0.1, $\pm$0.07, $\pm$0.05, $\pm$0.02, $\pm$0.01, 0.0, +0.15, +0.25, +0.35\}. 
This is a subset of the values used by L16 and is sufficient for our purposes since we are only interested in linear and quadratic bias coefficients. 

Initial conditions (ICs) for each simulation were generated using $2^{\rm nd}$ order Lagrangian perturbation theory using the code \textsc{music} \citep{hahn11-music}\footnote{http://www.phys.ethz.ch/$\sim$hahn/MUSIC/index.html}. We start the simulations at a cosmic time corresponding to $z=49$ in the fiducial cosmology; we note that this implies different starting redshifts for simulations corresponding to different $\dL$. The shape of the transfer function for all cosmologies is identical, and is obtained using the code \textsc{camb} \citep{camb}\footnote{http://camb.info}. 
As discussed by \citet{wsck15a}, the only difference in transfer functions across different cosmologies is due to the growth factor. Since \textsc{music} uses $\sigma_8$ for normalising the ICs, however, we must include an additional factor that accounts for the change in the Hubble parameter $h$ across cosmologies. We describe our method in detail in Appendix~\ref{app:ICs}. We use the same random number seed for the ICs of all the simulations in a single set of Separate Universe cosmologies. As noted also by L16, this largely cancels the sample variance of our results. To further reduce the error bars on our bias estimators, we have run \NSIM\ simulation sets by changing the seed.

For finding haloes, we use the code \textsc{rockstar} \citep{behroozi13-rockstar}\footnote{http://code.google.com/p/rockstar/}, which uses a Friends-of-Friends (FoF) algorithm in $6$-dimensional phase space to locate halo centers, and then assigns spherical overdensity masses to the haloes. We use a mass definition corresponding to a radius $R_{\rm 200b}$ at which the enclosed density is $200$ times the \emph{fiducial} background density (see, e.g., L16); we refer to these masses as $m_{\rm 200b}$ below. Note that this implies a different density threshold at different redshifts even in the same simulation (except for the fiducial $\dL=0$ runs); this is conveniently handled by \textsc{rockstar} which allows for multiple user-specified mass definitions. 

Since we will be interested in the environment dependence of internal halo properties, in particular halo concentration, it is important to ensure that our results are not biased by substructure and/or numerical artefacts. To this end, we discard subhaloes from our analysis, as well as unrelaxed haloes for which the `virial' ratio of kinetic and potential energies $\eta\equiv2T/|U|$ satisfies $\eta\geq2$ (corresponding to unbound objects). The subhalo criterion removes $\sim10\%$ ($\sim7\%$) of all objects at $z=0$ ($z=1$) in the fiducial cosmology, while the virial ratio cut removes an additional $\sim4\%$ ($\sim11\%$) of objects. The discarded objects are typically of low mass \citep[see, e.g.,][for a detailed discussion]{Bett+07}.

\begin{figure*}
\centering
\includegraphics[width=0.9\textwidth]{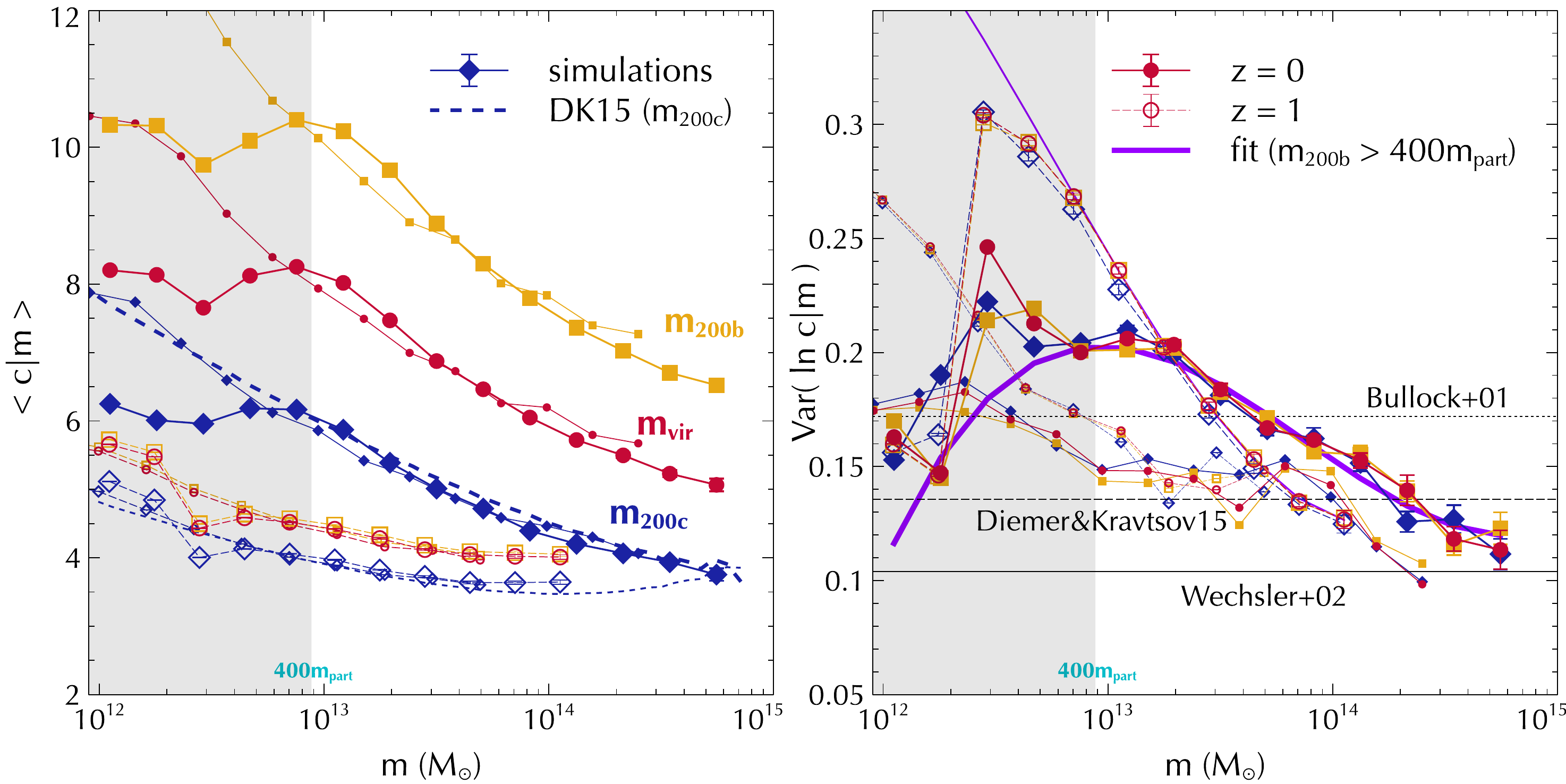}
\caption{\emph{(Left panel):} Median concentration in bins of halo mass averaged over \NSIM\ realisations of our $512^3$-particle simulation with fiducial cosmology at $z=0$ (large filled symbols) and $z=1$ (large empty symbols) for various mass definitions: $m_{\rm 200b}$ (yellow squares), $m_{\rm vir}$ (red circles) and $m_{\rm 200c}$ (blue diamonds). The smaller symbols show corresponding measurements in the $1024^3$-particle simulation. Smooth blue curves show the fitting function from \citet[][DK15]{dk15} at $z=0$ (dashed) and $z=1$ (dotted) for the $m_{\rm 200c}$-haloes. \emph{(Right panel):} Central $68.3\%$ scatter in $\ln c$ (equivalent of the square of one Gaussian standard deviation) for the same halo populations as in the left panel and labelled identically. Horizontal lines indicate the constant values reported by \citet{bullock+01}, \citet{wechsler+02} and DK15, as labelled. Solid purple curves show cubic polynomial fits to the $m_{\rm 200b}$ measurements at $z=0$ (thick) and $z=1$ (thin) as a function of $\log(m)$, constrained using measurements in the range $m_{\rm 200b} > 400m_{\rm part}$. The shaded region in each panel indicates the masses to be discarded in the Separate Universe analysis. See text for details.}
\label{fig:cmz}
\end{figure*}

In addition to these simulations, we have generated a single realisation of the fiducial cosmology for the same box size but containing $1024^3$ particles, using a $2048^3$ PM grid and force resolution $\epsilon=3.5$kpc comoving. We use this high resolution simulation to test for convergence of the halo mass function and halo concentration distribution in the fiducial cosmology, and use the $512^3$-particle runs as the default simulations for our analysis.

All the calculations required for generating the configuration files for \textsc{music}, \textsc{gadget-2} and \textsc{rockstar} were performed using NumPy \citep{vanderwalt-numpy}\footnote{http://www.numpy.org}. We have made our Python script publicly available\footnote{https://bitbucket.org/aparanjape/separateuniversescripts}, along with a Shell script that generates the three configuration files for any choice of fiducial $\Lambda$CDM cosmology and linear density contrast \dL. As a check, the symbols in Figure~\ref{fig:mf} show the mass function of $m_{\rm 200b}$-haloes that survive the cuts mentioned above in the fiducial $\dL=0$ simulation at two redshifts (averaged over \NSIM\ realisations), compared to the analytical fit from \citet[][hereafter, T08]{Tinker08}. The continuous lines show the corresponding measurements in our high-resolution box. The shaded region indicates haloes with $m_{\rm 200b} < 400m_{\rm part}$ which we discard in our analysis of the Separate Universe simulations. (Throughout, we use $m_{\rm part}$ to denote the particle mass in our default, $512^3$-particle simulations.) We discuss the choice of this threshold below.

\subsection{Concentration-mass-redshift relation in the fiducial cosmology}
\label{subsec:cmz_fid}
\noindent
The analysis below relies heavily on the distributions of halo concentration at fixed mass and redshift, so that it is important to ensure that these are measured in an unbiased manner. In practice, we use the values of halo scale radius $r_{\rm s}$ reported by \textsc{rockstar}, which are derived from fitting NFW profiles \citep*{nfw97} to the dark matter density measured in concentric shells around the halo centers-of-mass \citep{behroozi13-rockstar}. The halo concentration $c$ is then $c=c_{\rm 200b}\equiv R_{\rm 200b}/r_{\rm s}$. (Similar definitions hold for other mass definitions $m_{\rm 200b}\to m_{\Delta}$, with $c\to c_\Delta \equiv R_\Delta/r_{\rm s}$.) 

The dependence of the median values and scatter of $c$ on halo mass and redshift has been thoroughly explored in the literature. The median $\bar c(m,z)$ is known to be an approximately universal function of `peak height'
$\nu(m,z)$ \citep{ludlow+14,dk15}.
\footnote{The peak height is defined by
  $\nu(m,z) \equiv \delta_{\rm c}(z) / \sigma_0(m)$, where $\delta_{\rm c}(z)$ is the critical threshold for spherical
  collapse and $\sigma_0(m)$ is the standard deviation of linear fluctuations
  smoothed on comoving scale $R=(3m/4\pi\bar\rho)^{1/3}$ with a tophat window,
  both quantities being extrapolated to $z=0$ using linear theory.}
The scatter in $c$, after correcting for Poisson and fitting errors, is approximately \emph{independent} of mass and redshift \citep{bullock+01,wechsler+02}. The shape of the distribution of $c$ is approximately Lognormal \citep{wechsler+02,st04}, with noticeable non-Gaussian tails \citep{bullock+01,wechsler+02,dk15}.

The left panel of Figure~\ref{fig:cmz} shows the median concentration-mass relation at two redshifts, for various mass definitions, averaged over \NSIM\ realisations of our default simulation with the \emph{fiducial} cosmology (large symbols). In each case, we see a distinct departure from monotonicity at $m\lesssim 400m_{\rm part}$. Since this feature is independent of mass definition (and also approximately independent of redshift) we interpret it as a consequence of mass resolution. This is supported by the fact that the corresponding measurements in the high resolution box (shown as the smaller symbols) remain monotonic until much smaller masses, and also by the good agreement until $m_{\rm 200c}\sim400m_{\rm part}$ between our measurements for $m_{\rm 200c}$-haloes and the analytical fit of \citet[][dashed blue curves]{dk15}, which was based on measurements using \textsc{rockstar} haloes with very stringent requirements on force and mass resolution. This is also consistent with the departure of the measured mass fraction from the T08 fit at $m_{\rm 200b}\lesssim400m_{\rm part}$ in the right panel of Figure~\ref{fig:mf}, at both redshifts, and the fact the high resolution measurements in that Figure are in better agreement with the T08 fit at these low masses. 
We therefore use $m_{\rm min} = 400m_{\rm part}$ as the minimum mass for all our subsequent analysis.

The right panel of Figure~\ref{fig:cmz} shows the central $68.3\%$ scatter (i.e., the equivalent to the square of one Gaussian standard deviation), for the same halo populations as in the left panel. Although this scatter is clearly independent of choice of mass definition, we do see a substantial trend with mass and redshift in the default simulations. This trend is significantly suppressed in the high resolution simulation. Previous authors have found values of scatter that are nearly constant with mass and redshift \citep[the horizontal lines indicate the values reported by][as labelled]{bullock+01,wechsler+02,dk15}. We interpret the trends in our measurements as arising mainly due to resolution-dependent NFW fitting errors; these are consistent with the trends seen between the uncorrected and corrected scatter in, e.g., Figure~4 of \citet{bullock+01}. We calibrate these trends by fitting cubic polynomials in $\log(m)$ to the measurements at each redshift, using bins with $m > 400m_{\rm part}$. The thick and thin solid purple curves in the Figure show these fits for the $m_{\rm 200b}$-haloes. We will use these fits below to statistically correct our measurements of the concentration distributions in all our Separate Universe simulations.

\section{Separate Universe assembly bias}
\label{sec:biasSU}
\noindent
We now turn to the analysis of the Separate Universe simulations and the measurement of halo assembly bias. We start by  recapitulating the L16 technique for measuring non-linear halo bias as a function of halo mass. This will help define our notation and also serve as a sanity check on our simulations. Following this, we present the analysis of assembly bias in the same simulations using two different techniques.

\subsection{Average bias coefficients}
\label{subsec:avgbias}
\noindent
Denoting the differential number density of haloes with mass in the range $(m,m+\der m)$ in a simulation with initial overdensity \dL\ as $n(m|\dL)$, the local Lagrangian bias coefficients $b_{n}^{\rm L}$ satisfy the relation
\begin{align}
\dHL(m,\dL) 
&\equiv \frac{n(m|\dL)}{n(m|\dL=0)} - 1 \notag\\
&= \sum_{n=1}^\infty\,\frac{b_{n}^{\rm L}(m)}{n!}\,\dL^n\,,
\label{eq:localLagbias}
\end{align}
where we suppressed the redshift dependence. We follow L16 and fit $4^{\rm th}$ order polynomials\footnote{As discussed by L16, a rule of thumb is to fit a polynomial of degree $n+2$ for reporting bias of order $n$. We have checked that fitting $5^{\rm th}$ order polynomials does not significantly alter our results.} to the measurements of $\dHL(\dL,m)$ in each mass bin.  We use the mean value and standard error of $\dHL(\dL,m)$ over \NSIM\ realisations for the polynomial fits. The errors on the reported coefficient values then correspond to the diagonal entries of the covariance matrix recovered from the fit.

\begin{figure}
\centering
\includegraphics[width=0.45\textwidth]{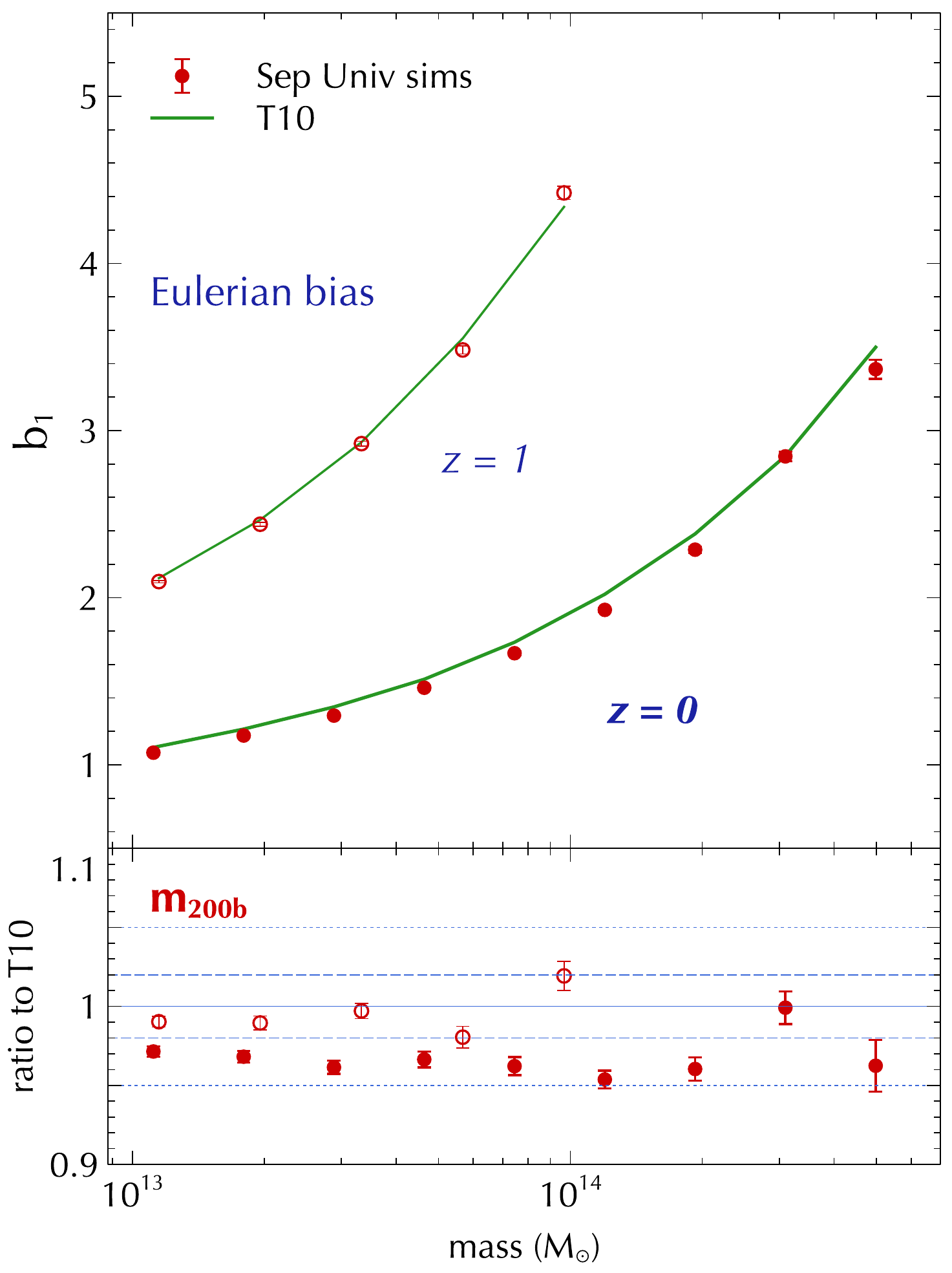}
\caption{\emph{(Top panel):} Linear Eulerian bias coefficient measured using \NSIM\ realisations of the Separate Universe simulations (points with error bars) and the analytical fit from \citet[][T10, solid green curves]{Tinker10}. \emph{(Bottom panel):} Ratio of the measurements with the T10 fit. In both panels, filled (empty) symbols show results at $z=0$ ($z=1$). Error bars are derived as described in the text.}
\label{fig:b1-SU}
\end{figure}

The corresponding Eulerian parameters $b_n$ follow from the relation $1+\dH = (1+\dHL)(1+\delta) = 1+\sum_{n=1}^\infty(b_n/n!)\delta^n$ \citep{mw96} where the nonlinear $\delta(\dL)$ can be approximated using spherical evolution as $\delta = \dL\,g(z) + (17/21)\dL^2\,g(z)^2 + \Cal{O}(\dL^3)$ \citep{Bernardeau92,wsck15b}, where $g(z)\equiv D(z)/D(0)$ with $D(z)$ the linear theory growth factor of the fiducial cosmology, leading to
\begin{align}
b_1 &= 1 + b_1^{\rm L}\,g(z)^{-1}\,, \notag\\
b_2 &= b_2^{\rm L}\,g(z)^{-2} + \frac{8}{21}\,b_1^{\rm L}\,g(z)^{-1}\,.
\label{eq:localEulbias}
\end{align}
Note that the combinations $b_n^{\rm L}\,g(z)^{-n}$ correspond to Lagrangian bias coefficients with respect to the linear density extrapolated to the measurement redshift.
The top panel of Figure~\ref{fig:b1-SU} shows the measured $b_1$ at two redshifts (points with error bars) and the corresponding analytical expression from \citet[][heareafter, T10]{Tinker10} as the solid curves. The bottom panel shows the ratio between the measurements and the T10 fit\footnote{To ensure a fair comparison, we binned the analytical T10 results over the same mass bins used for the measurements, applying a number weighting using the T08 mass function.}. The fit describes our measurements to within $\sim5\%$ ($\sim2\%$) over the range of masses explored at $z=0$ ($z=1)$.

\begin{figure}
\centering
\includegraphics[width=0.45\textwidth]{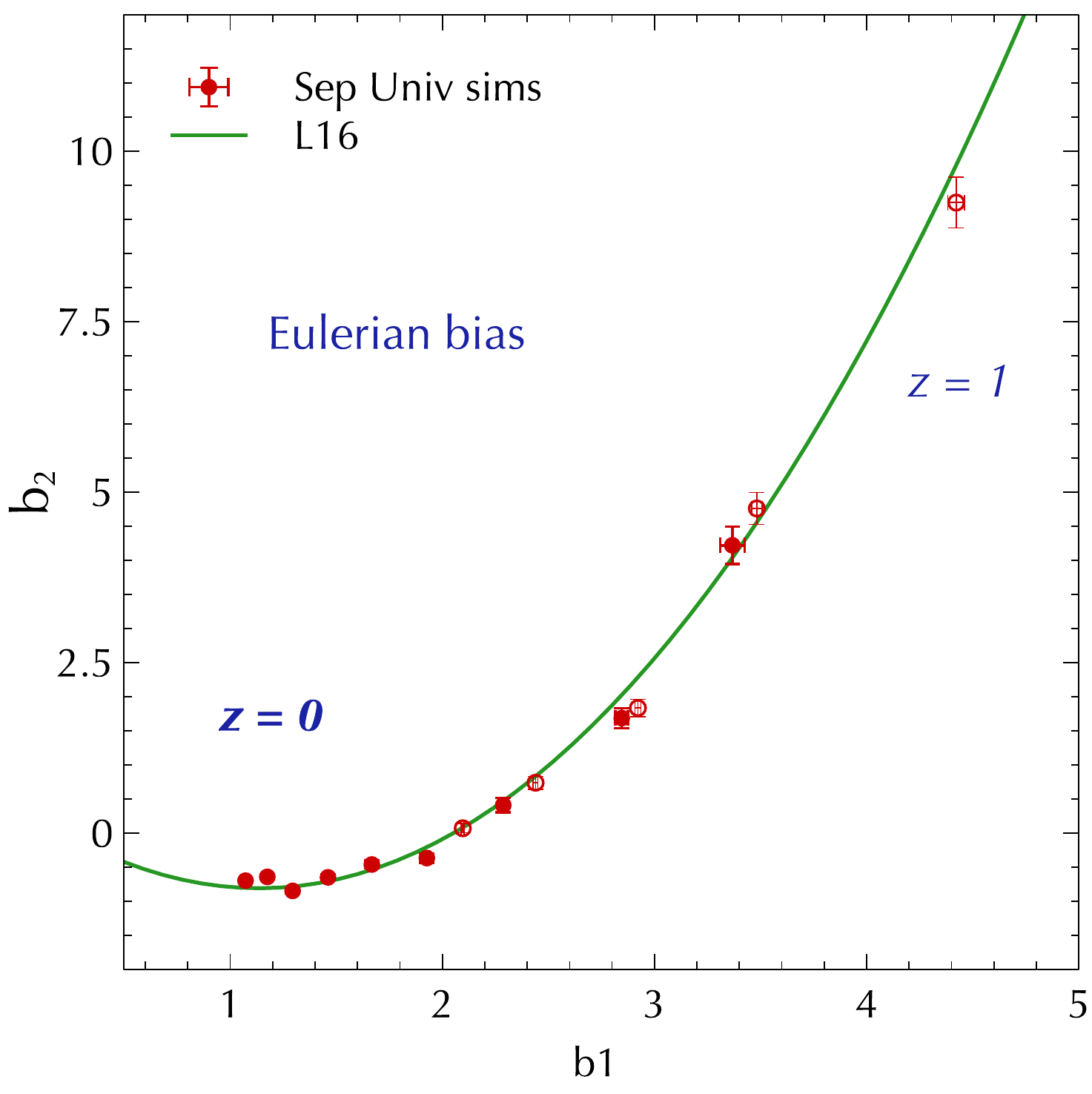}
\caption{Quadratic bias $b_2$ as a function of the linear $b_1$, measured using \NSIM\ realisations of the Separate Universe simulations. The solid curve shows the fitting function from equation 5.2 of \citet[][L16]{lwbs16}. The formatting is identical to that in Figure~\ref{fig:b1-SU}.}
\label{fig:b2-SU}
\end{figure}

Figure~\ref{fig:b2-SU} shows the measured $b_2$ as a function of $b_1$ (points with error bars), with the solid curve showing the fitting function from L16 (their equation 5.2). The measurements are clearly almost universal, with the fit describing both the $z=0$ and $z=1$ data. Figures~\ref{fig:b1-SU} and~\ref{fig:b2-SU} thus confirm the results reported by L16 and show that our basic setup for measuring bias from Separate Universe simulations is working correctly.


\subsection{Assembly bias from binning in concentration}
\label{subsec:concbin}
\noindent
The L16 method above can be generalised in a straightforward manner to measure assembly bias. Essentially, we use the fact that one of the most prominent signatures of this effect is a difference in large scale halo bias between haloes of the same mass but different concentrations. Physically, this is closely related to the fact that halo concentrations are tightly correlated with halo formation history, and the latter is correlated with halo environment at fixed halo mass. We exploit this concentration dependence by simply binning halo counts in our Separate Universe simulations in both mass \emph{and} concentration.

Since halo concentrations are known to be close to Lognormally distributed (see also below), it becomes convenient to  define a standardized variable $s$ using
\be
s \equiv \ln\left(c/\bar c\right) / \sigma_0\,,
\label{eq:sdef}
\ee
where
\begin{align}
\ln \bar c &\equiv \avg{\ln c|m,\dL=0}\,,\notag\\
\sigma_0^2 &\equiv {\rm Var}\left(\ln c|m,\dL=0\right)\,,
\label{eq:cbarsig0def}
\end{align}
are the median and central $68.3\%$ scatter of the concentrations in the fiducial simulation. We use this definition of $s$ for \emph{all} the simulations, the reason for which will become apparent below\footnote{Throughout this work, we will use analytical results that assume a Gaussian distribution for the log-concentration. To reduce the effects of systematics associated with the non-Gaussianity of the measured distributions, we will always use the median and central $68.3\%$ scatter instead of the mean and variance, respectively, since the former are less sensitive to non-Gaussian tails and outliers. On the theoretical side, since we always ignore the non-Gaussianity, the median and $68.3\%$ scatter are equivalent to the mean and variance, so we will use notation appropriate for the latter two.}. As discussed previously, however, the measured scatter in the fiducial simulation (and, indeed in all our Separate Universe simulations) has a substantial mass dependence due to fitting errors. We statistically correct for this mass dependence as described in Appendix~\ref{app:NFWfitcorr}.

By binning haloes simultaneously in mass as well as concentration (or $s$), we can define a Lagrangian halo overdensity $\dHL(m,s,\dL)$. Taylor expanding this in powers of \dL\ then gives us Lagrangian bias parameters that depend on both mass and concentration,
\begin{align}
\dHL(m,s,\dL) 
&\equiv \frac{n(m,s|\dL)}{n(m,s|\dL=0)} - 1 \notag\\
&= \sum_{n=1}^\infty\,\frac{b_{n}^{\rm L}(m,s)}{n!}\,\dL^n\,,
\label{eq:localLagAB}
\end{align}
where $n(m,s|\dL)$ is the differential number density of halos with mass in the range $(m,m+\der m)$ and standardized concentration in the range $(s,s+\der s)$, in a Separate Universe simulation with overdensity \dL.

\begin{figure}
\centering
\includegraphics[width=0.45\textwidth]{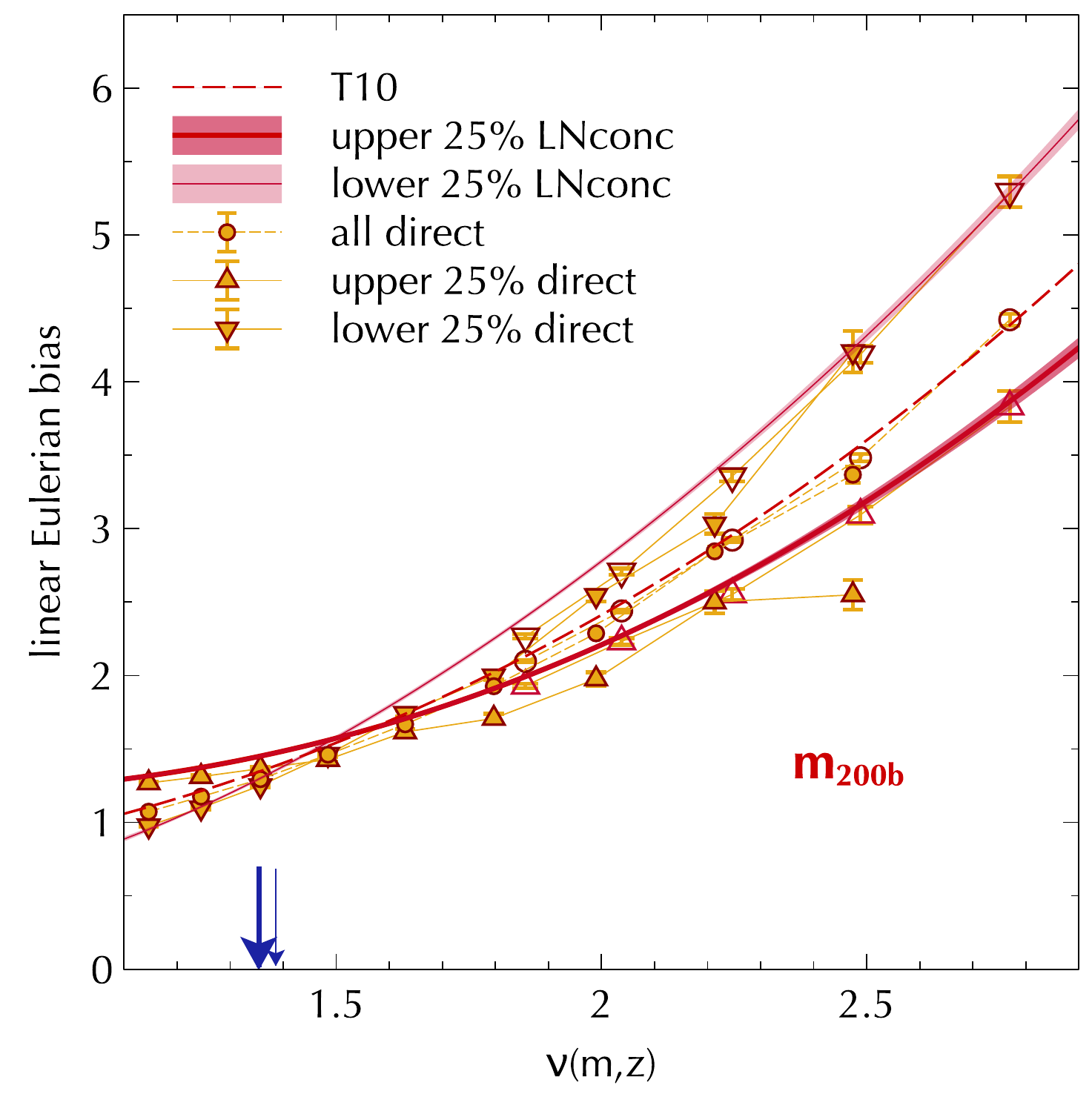}
\caption{Halo bias from Separate Universe simulations as a function of peak height $\nu$, obtained by directly binning haloes in mass and concentration (points with errors, see section~\ref{subsec:concbin} for details) and from the Lognormal-based method described in section~\ref{subsec:lognormal}. For the binned technique, results are shown for all haloes (circles) and haloes with concentrations satisfying $s > 0.675$ (triangles) and $s < -0.675$ (inverted triangles); the latter two approximately correspond to the upper and lower quartiles of concentration in the fiducial cosmology, respectively. Filled (empty) symbols show results at $z=0$ ($z=1$). For the Lognormal-based method, the dashed curve shows the T10 fit for all haloes, while the thick (thin) solid curves show the result of averaging $b_1(m,s)$ (equations~\ref{eq:b1Lag(m,s)} and~\ref{eq:localEulbias}) over $s > 0.675$ ($s < -0.675$). The bands around these curves show $1\sigma$ errors as described in the text. The thick (thin) arrow indicates the value $\nu=\nu_\ast(z)$ corresponding to the characteristic mass $m_\ast(z)$ at $z=0$ ($z=1$) as described in the caption of Figure~\ref{fig:mf}.} 
\label{fig:abSUfitVsdirect}
\end{figure}

The points with errors in Figure~\ref{fig:abSUfitVsdirect} show the resulting linear Eulerian bias (see equation~\ref{eq:localEulbias}) for haloes in the upper and lower quartiles of concentration of the fiducial cosmology (triangles), and for all haloes (circles). The bias coefficient in each mass and concentration bin was estimated using the L16 procedure: we fit $4^{\rm th}$ order polynomials in \dL\ to the relation \eqref{eq:localLagAB} and report the linear coefficient and its error from the least squares fit. To emphasize the near-universality of the measurements, we show results as a function of peak height $\nu(m,z)$. The arrows indicate the values of $\nu$ corresponding to the characteristic mass scales $m_\ast(z)$ as in Figure~\ref{fig:mf}; the small difference between the values at different redshifts (which we denote $\nu_\ast(z)$) reflects the small amount of non-universality in the T08 mass function.

Assembly bias is clearly evident in the Figure, with the bias of lower quartile haloes being higher than that of upper quartile haloes at large $\nu$, and this trend reversing at small $\nu$. We see that this inversion occurs close to the characteristic scale $\nu_\ast$. Interestingly, previous analyses of assembly bias have obtained a similar inversion at much lower masses, corresponding to $\nu\simeq1$; we discuss possible reasons for this discrepancy in section~\ref{sec:discuss}.

In principle, we could now characterise assembly bias by fitting these binned results as a function of $s$ and $\nu$ \citep[see, e.g.,][]{wechsler+06}. There are two disadvantages to this approach, however. Firstly, by binning in $s$ we necessarily increase the noise in the assembly bias measurements as compared to that in the all-halo bias. Secondly, the fitting function cannot be arbitrary, since it must reproduce the correct all-halo bias upon averaging over concentrations. This introduces additional complexity in the fitting procedure. In the next section we show how these problems can be circumvented by exploiting the near-Lognormal shape of the concentration distribution.

\subsection{Assembly bias assuming Lognormal concentrations}
\label{subsec:lognormal}
\noindent
In this section we show how assembly bias can be easily formulated in terms of the distributions $p(s|m,\dL)$ of the variable $s$ (equation~\ref{eq:sdef}) at fixed halo mass, in Separate Universe simulations with overdensity \dL. We then use the near-Lognormal nature of the concentration distribution at fixed halo mass to obtain convenient analytical expressions for $b_1^{\rm L}(m,s)$ and $b_2^{\rm L}(m,s)$.

In general, noting that
\be
n(m,s|\dL) = n(m|\dL)\,p(s|m,\dL)\,, 
\label{eq:n(m,s|dL)}
\ee
the ratio $n(m,s|\dL)/n(m,s|\dL=0)$ in \eqn{eq:localLagAB} separates into the product 
\begin{align}
1+\dHL(m,s,\dL) &= \frac{n(m,s|\dL)}{n(m,s|\dL=0)}\notag\\ 
&=  \left(1+\dHL(m,\dL)\right) \times \left[\frac{p(s|m,\dL)}{p(s|m,\dL=0)}\right]
\label{eq:concdepratio}
\end{align}
where $\dHL(m,\dL)$ is the concentration-independent halo overdensity from \eqn{eq:localLagbias}. 
Next, we expand the ratio in square brackets in powers of \dL\ by defining mass and concentration dependent Taylor coefficients $\Cal{P}^{\rm L}_n(s,m)$ using
\be
\frac{p(s|m,\dL)}{p(s|m,\dL=0)} \equiv 1 + \sum_{n=1}^\infty\,\frac{\Cal{P}_{n}^{\rm L}(m,s)}{n!}\,\dL^n\,.
\label{eq:p(s|d)taylor}
\ee
Combining \eqns{eq:localLagbias}, \eqref{eq:localLagAB}, \eqref{eq:concdepratio} and~\eqref{eq:p(s|d)taylor}, we can then easily prove the identity
\begin{align}
b_n^{\rm L}(m,s) &= b_n^{\rm L}(m) + \Cal{P}_n^{\rm L}(m,s) \notag\\
&\ph{b_n^{\rm L}(m)}
+ \sum_{k=1}^{n-1}\binom{n}{k}\,b_k^{\rm L}(m)\Cal{P}_{n-k}^{\rm L}(m,s)\,,
\label{eq:bnLag(m,s)}
\end{align}
so that the concentration dependence of the bias coefficients is completely determined by the relative shapes of the distributions $p(s|m,\dL)$ and $p(s|m,\dL=0)$. 

Moreover, it is clear from \eqn{eq:p(s|d)taylor} that, in order to ensure the normalisation $\int\der s\,p(s|m,\dL)=1$, we must have
\be
\avg{\Cal{P}_n^{\rm L}(m,s)} \equiv \int_{-\infty}^{\infty}\der s\,p(s|m,\dL=0)\,\Cal{P}_n^{\rm L}(m,s) = 0\,,
\label{eq:Pn-avg}
\ee
for \emph{all} $n\geq 1$. Averaging over concentrations in \eqn{eq:bnLag(m,s)} then immediately gives us
\be
\avg{b_n^{\rm L}(m,s)} \equiv \int_{-\infty}^\infty\der s\,p(s|m,\dL=0)\,b_n^{\rm L}(m,s) = b_n^{\rm L}(m)\,,
\label{eq:<bn|m,s>}
\ee
for $n\geq1$. This is also clear upon noting that the parameters $b_n^{\rm L}(m)$ are Taylor coefficients of $1+\dHL(m,\dL) = \int\der s\,n(m,s|\dL) / \int\der s\,n(m,s|\dL=0)$, and writing the integrand of the numerator in terms of the parameters $b_n^{\rm L}(m,s)$. In other words, our formulation \emph{guarantees} that the concentration-dependent bias parameters correctly and self-consistently average to the concentration-independent ones. We emphasize that the analysis above does not assume any specific shape for the fiducial distribution $p(s|m,\dL=0)$. 

\begin{figure}
\centering
\includegraphics[width=0.48\textwidth]{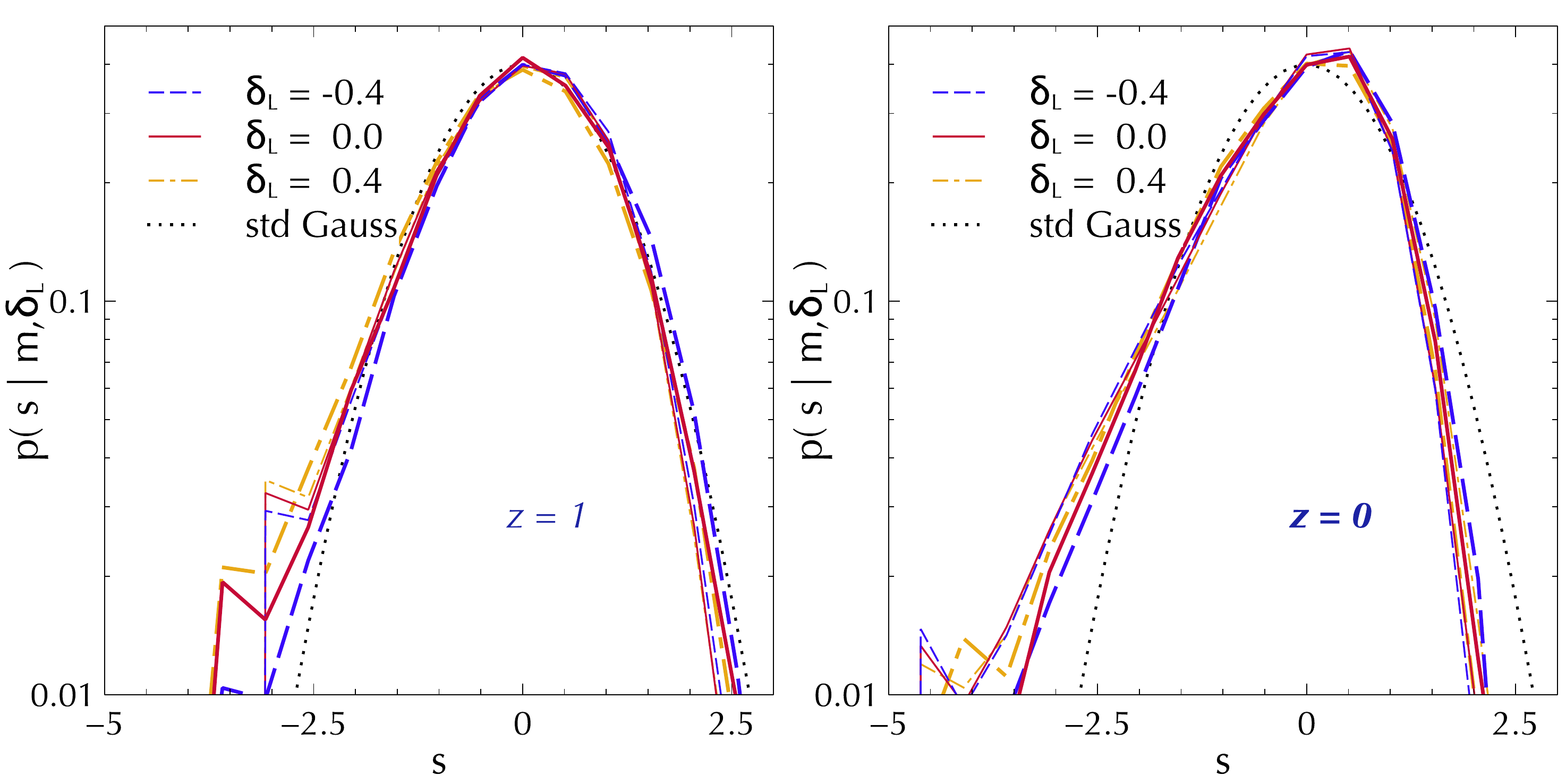}
\caption{Distribution $p(s|m,\dL)$ of standardized concentration $s$ (equation~\ref{eq:sdef}) at fixed halo mass $m$ and overdensity \dL, at $z=1$ \emph{(left panel)} and  $z=0$ \emph{(right panel)}, averaged over \NSIM\ realisations. The solid red, dashed blue and dot-dashed yellow curves show results for $\dL=0.0,-0.4,0.4$, respectively. In the \emph{left panel}, the thick (thin) curves are for $m=10^{13.75} (10^{13.1})\Msun$, while in the \emph{right panel}, the thick (thin) curves are for $m = 10^{14.3} (10^{13.0})\Msun$. Dotted black curve in each panel shows the standard Gaussian distribution $p(s)={\rm e}^{-s^2/2}/\sqrt{2\pi}$. The $s$ value for each halo was not corrected for NFW fitting errors in making these histograms. See text for a discussion.} 
\label{fig:s_hist}
\end{figure}

To proceed further, we must make some assumptions about $p(s|m,\dL)$. The near-Lognormal nature of the concentration distribution at fixed halo mass in the fiducial simulation means that the distribution $p(s|m,\dL=0)$ should be close to a standard Gaussian (zero mean and unit variance), independent of halo mass. 
Figure~\ref{fig:s_hist} shows the distribution $p(s|m,\dL)$ for various choices of mass and overdensity at two redshifts, averaged over \NSIM\ realisations. At each redshift, we see that there is only a weak mass dependence (compare the thick and thin curves for each line style). There is, however, a monotonic dependence of the distribution on \dL, which will be important below. Comparing with the standard Gaussian distribution (dotted black curve in each panel), we see that $p(s|m,\dL=0)$ is indeed quite close to being Gaussian at $z=1$, but has a noticeable skewness towards small concentrations by $z=0$. We will comment on the effects of ignoring this non-Gaussianity below.

We also see in the Figure that these aspects of the non-Gaussianity of $s$, as well as the details of the skewness, are broadly independent of \dL. In other words, the distributions $p(s|m,\dL)$ in simulations with $\dL\neq0$ do not pick up any additional substantial non-Gaussianity, implying that these are also reasonably well-modelled as \emph{Gaussians} with mean $\mu(m,\dL)$ and variance $\sigma^2(m,\dL)$:
\be
\mu(m,\dL) \equiv \avg{s|m,\dL}\,;\quad \sigma^2(m,\dL) \equiv {\rm Var}\left(s|m,\dL\right)\,.
\label{eq:musigmadef}
\ee
By construction, we have $\mu(m,\dL=0)=0$ and $\sigma^2(m,\dL=0)=1$, so that it is useful to define Taylor expansion coefficients $\mu^{\rm L}_n(m)$ and $\Sigma^{\rm L}_n(m)$ using
\begin{align}
\mu(m,\dL) &\equiv \sum_{n=1}^\infty \frac{\mu^{\rm L}_n(m)}{n!}\,\dL^n\,, \notag\\ 
\sigma^2(m,\dL) &\equiv 1 + \sum_{n=1}^\infty \frac{\Sigma^{\rm L}_n(m)}{n!}\,\dL^n\,.
\label{eq:munSigndef}
\end{align}
Taking $p(s|m,\dL=0)$ to be a standard Gaussian distribution, we can then explicitly evaluate the coefficients $\Cal{P}_n^{\rm L}(m,s)$ in terms of Hermite polynomials (which all average to zero over the standard Gaussian distribution) and the coefficients $\mu^{\rm L}_n(m)$ and $\Sigma^{\rm L}_n(m)$. For $n=1,2$ this leads to
\begin{align}
b_1^{\rm L}(m,s) 
&= b_1^{\rm L}(m) + \mu^{\rm L}_1(m)\,H_1(s) + \frac12\Sigma^{\rm L}_1(m)\,H_2(s)\,,
\label{eq:b1Lag(m,s)}\\
b_2^{\rm L}(m,s) 
&= b_2^{\rm L}(m) + \left\{\mu^{\rm L}_2(m)+2b_1^{\rm L}(m)\mu^{\rm L}_1(m)\right\}\,H_1(s)\notag\\
&\ph{b_2^{\rm L}}
+ \left\{\mu^{\rm L}_1(m)^2 + b_1^{\rm L}(m)\Sigma^{\rm L}_1(m) + \frac12\Sigma^{\rm L}_2(m)\right\}\,H_2(s)\notag\\
&\ph{b_2^{\rm L}(m)}
+\mu^{\rm L}_1(m)\,\Sigma^{\rm L}_1(m)\,H_3(s) + \frac14\Sigma^{\rm L}_1(m)^2\,H_4(s)\,,
\label{eq:b2Lag(m,s)}
\end{align}
where the $H_n(s)$ are the `probabilist's' Hermite polynomials defined by $H_n(s)\equiv \e{s^2/2}(-\der/\der s)^n\e{-s^2/2}$, so that
\begin{align}
&H_1(s) = s~ ;\qquad\quad~~ H_2(s) = s^2-1\,,\notag\\
&H_3(s) = s^3-3s~ ;\quad H_4(s) = s^4-6s^2+3\,.
\label{eq:Hermite-1to4}
\end{align}
We refer the reader to Appendix~\ref{app:LNcalc} for the details of the above derivation.
Finally, the spurious mass dependence introduced by fitting errors (see section~\ref{subsec:cmz_fid}) affects the measurements of $\mu(m,\dL)$ and $\sigma^2(m,\dL)$. In Appendix~\ref{app:NFWfitcorr}, we show how this dependence can be removed using measurements of the mass dependence of scatter in the fiducial cosmology (shown in the right panel of Figure~\ref{fig:cmz}).

\begin{figure*}
\centering
\includegraphics[width=0.9\textwidth]{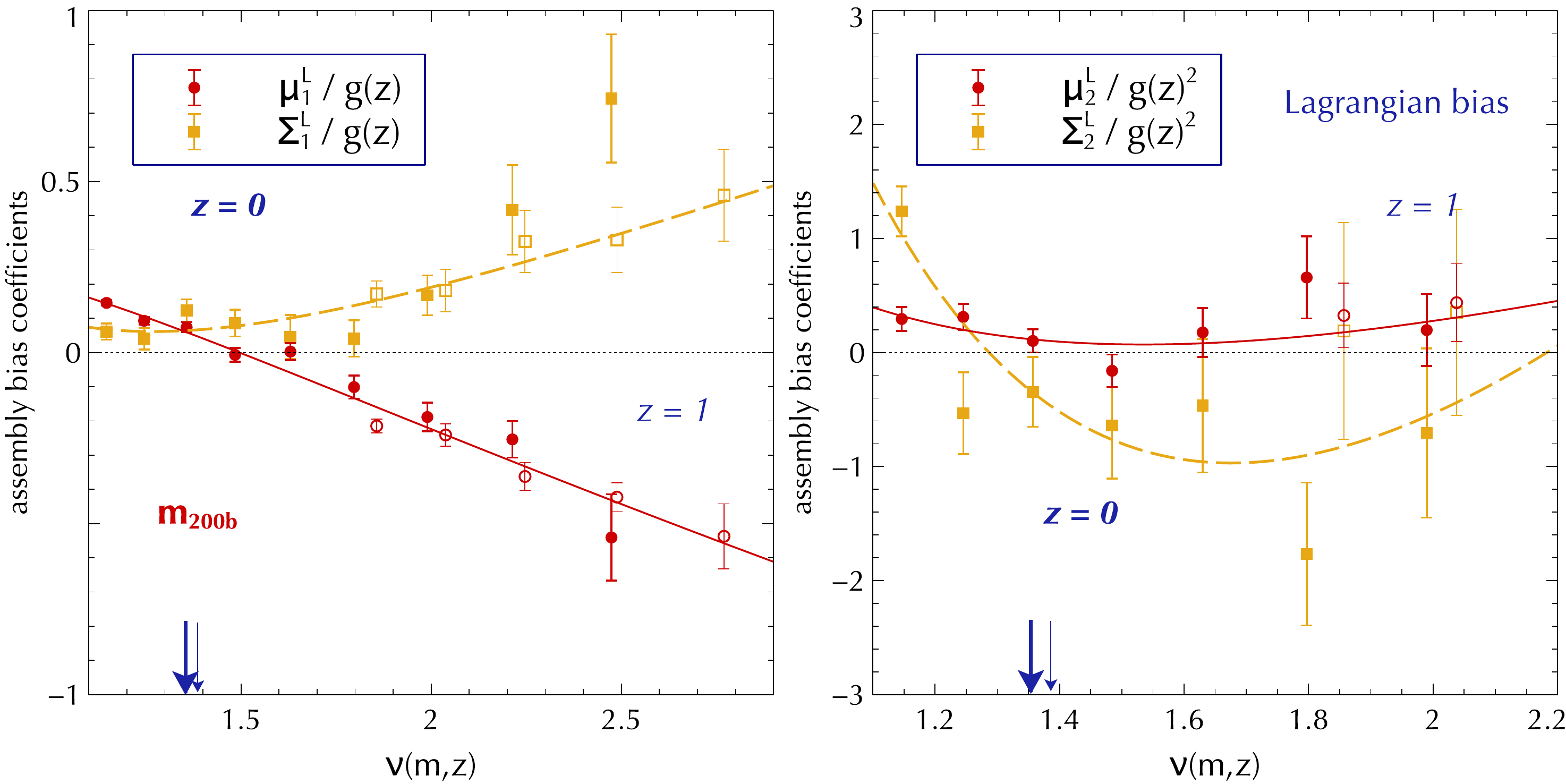}
\caption{Lagrangian assembly bias coefficients $\mu_n^{\rm L}$ and $\Sigma_n^{\rm L}$ (see equations~\ref{eq:munSigndef}-\ref{eq:b2Lag(m,s)}), extrapolated to the measurement redshift by dividing by $g(z)^{n}$ and shown as functions of $\nu(m_{\rm 200b},z)$. The left panel shows the linear coefficients and the right panel shows the quadratic coefficients. Filled (empty) symbols with error bars show measurements using the Separate Universe technique at $z=0$ ($z=1$), with red circles (yellow squares) showing $\mu_n^{\rm L}$ ($\Sigma_n^{\rm L}$). The corresponding smooth red solid (yellow dashed) curves show the best fit polynomials for $\mu_n^{\rm L}$ ($\Sigma_n^{\rm L}$) in the variable $y\equiv\log_{10}(\nu/1.5)$, with coefficients and errors reported in Table~\ref{tab:ABcoeffs}. The arrows indicate the values of $\nu_\ast(z)$ as in Figure~\ref{fig:abSUfitVsdirect}. See text for a discussion.} 
\label{fig:abSUcoeffs}
\end{figure*}

This formulation then lends itself to a straightforward and convenient extension of the L16 method, yielding an accurate characterisation of the concentration dependence of halo bias. The procedure is as follows:
\begin{itemize}
\item {\bf Fit} the concentration-independent $b_1^{\rm L}(m)$, $b_2^{\rm L}(m)$ exactly as in L16.
\item For the $\dL=0$ simulation, {\bf measure} $\bar c(m)$ and $\sigma_0^2(m)$ (equation~\ref{eq:cbarsig0def}) in mass bins.
\item For each $\dL\neq0$ and every halo, use the measured halo concentrations $c$ to {\bf compute} the variable $s$ using \eqn{eq:sdef}.
\item For each \dL\ and every mass bin, {\bf measure} the median $\mu(m,\dL)$ and central $68.3\%$ scatter $\sigma^2(m,\dL)$ of the distribution $p(s|m,\dL)$.
\item {\bf Fit} the parameters $\mu^{\rm L}_1(m)$, $\mu^{\rm L}_2(m)$ and $\Sigma^{\rm L}_1(m)$, $\Sigma^{\rm L}_2(m)$ using the same approach as used for $b_1^{\rm L}(m)$, $b_2^{\rm L}(m)$.
\item {\bf Correct} the parameters $\mu^{\rm L}_n(m)$, $\Sigma^{\rm L}_n(m)$ as described in Appendix~\ref{app:NFWfitcorr}. 
\item The concentration dependent bias coefficients $b_n^{\rm L}(m,s)$ are then given by \eqns{eq:b1Lag(m,s)}-\eqref{eq:b2Lag(m,s)}.
\end{itemize}

We see that, compared to the L16 approach, we must perform two more polynomial fits for extracting the assembly bias coefficients $\mu_n^{\rm L}$ and $\Sigma_n^{\rm L}$, with the rest of the operations being direct measurements on the halo catalogs\footnote{The exception is the correction for the spurious mass and redshift dependence of $\sigma_0^2$, which involves additional fitting in the fiducial simulation.}. Additionally, the procedure above does not require binning in concentration, since we exploit the near-Gaussianity of $s$. This means we can use all haloes in a given mass bin in extracting the assembly bias coefficients, a distinct advantage over methods that split the halo population into bins of concentration (c.f. section~\ref{subsec:concbin}).

\begin{table*}
\begin{center}
\begin{tabular}{ccccccccccc}
\hline \hline
& $\mu_{10}$ & $\mu_{11}$ & $\mu_{12}$ & $\chi^2(10{\rm \,dof})$ &&& $\Sigma_{10}$ & $\Sigma_{11}$ & $\Sigma_{12}$ & $\chi^2(10{\rm \,dof})$\\[0.5 ex]
\hline
value & -0.0020 & -1.502 & -2.19 &  19.45  &&& 0.079 & 0.48 & 3.3 & 10.42 \\[0.5 ex]
std dev & 0.0092 & 0.059 & 0.63 &          &&& 0.019 & 0.14 & 1.3 & \\[0.5 ex]
\hline
corr $\mu_{10}$ & 1.0 & 0.483 & -0.780 &&& corr $\Sigma_{10}$ & 1.0 & 0.378 & -0.746 & \\[0.5 ex]
corr $\mu_{11}$ & -- & 1.0 & -0.264    &&& corr $\Sigma_{11}$ & -- & 1.0 & -0.393 & \\[0.5 ex]
\hline \hline
& $\mu_{20}$ & $\mu_{21}$ & $\mu_{22}$ & $\chi^2(10{\rm \,dof})$ &&& $\Sigma_{20}$ & $\Sigma_{21}$ & $\Sigma_{22}$  & $\chi^2(10{\rm \,dof})$ \\[0.5 ex]
\hline
value & 0.072 & -0.31 & 15.6 &  7.40   &&& -0.80 & -7.1 & 74 & 14.60 \\[0.5 ex]
std dev & 0.080 & 0.69 & 7.4 &         &&& 0.22 & 1.6 & 15 & \\[0.5 ex]
\hline
corr $\mu_{20}$ & 1.0 & 0.396 & -0.691 &&& corr $\Sigma_{20}$ & 1.0 & 0.468 & -0.742 & \\[0.5 ex]
corr $\mu_{21}$ & -- & 1.0 & -0.101     &&& corr $\Sigma_{21}$ & -- & 1.0 & -0.268 & \\[0.5 ex]
\hline
\end{tabular}
\end{center}
\caption{Best fit coefficients and covariance matrices of quadratic polynomial fits to $\mu_n^{\rm L}(y)$ and $\Sigma_n^{\rm L}(y)$ as functions of logarithmic peak height $y \equiv \log_{10}[\nu(m,z)/1.5]$, for $n=1,2$. 
The fits were performed in the range $1.1\leq\nu\leq2.9$, with coefficients organised as $\mu_1^{\rm L}(y)=\mu_{10} + \mu_{11}y + \mu_{12}y^2$, and so on. The upper and lower blocks give results for $n=1$ and $n=2$, respectively, with columns indicating the polynomial coefficients. In each block, the first row gives the least squares best fit value, the second row gives the standard deviation (square root of the diagonal of the covariance matrix), and the remaining rows give the correlation coefficients when paired with $\mu_{n\alpha}$ or $\Sigma_{n\alpha}$, where $\alpha$ counts over the polynomial coefficients.}
\label{tab:ABcoeffs}
\end{table*} 

For self-consistent results in the evolved field, we require a description of the concentration dependence of the Eulerian bias parameters $b_n(m,s)$. This can be obtained by the requirement
\be
1+\dH(m,s,\delta) = \left(1+\dHL(m,s,\dL)\right)\left(1+\delta\right)
\label{eq:dhEul(m,s,dL)}
\ee
which is formally identical to the relation averaged over concentration, since this follows solely from the requirement of mass conservation in the volume under consideration. In other words, the Eulerian $b_1(m,s)$ and $b_2(m,s)$ can be obtained using the concentration dependent Lagrangian parameters $b_n^{\rm L}(m,s)$ in the right hand sides of \eqns{eq:localEulbias}. 

\subsection{Results}
\label{subsec:results}
\noindent
The symbols with error bars in Figure~\ref{fig:abSUcoeffs} show the Lagrangian assembly bias coefficients $\{\mu_n^{\rm L},\Sigma_n^{\rm L}\}$ linearly extrapolated to the measurement redshift using the normalised growth factor $g(z)$, for $n=1,2$ at $z=0$ and $z=1$ as a function of peak height $\nu(m_{\rm 200b},z)$. The measurements used \NSIM\ realisations and were corrected using the procedure of Appendix~\ref{app:NFWfitcorr}. 
The arrows indicate the values of $\nu_\ast(z)$ corresponding to the characteristic mass scales $m_\ast(z)$ as in Figures~\ref{fig:mf} and~\ref{fig:abSUfitVsdirect}. Our choice of mass resolution and volume constrain us to the range $1.1 \leq \nu \leq 2.9$ for the linear coefficients. The measurement noise is larger for the quadratic coefficients, particularly at large masses; while we use these measurements in our analysis below, when showing plots we will restrict to the smaller range $1.1 \leq \nu \leq 2.2$ for these.

\begin{figure*}
\centering
\includegraphics[width=0.9\textwidth]{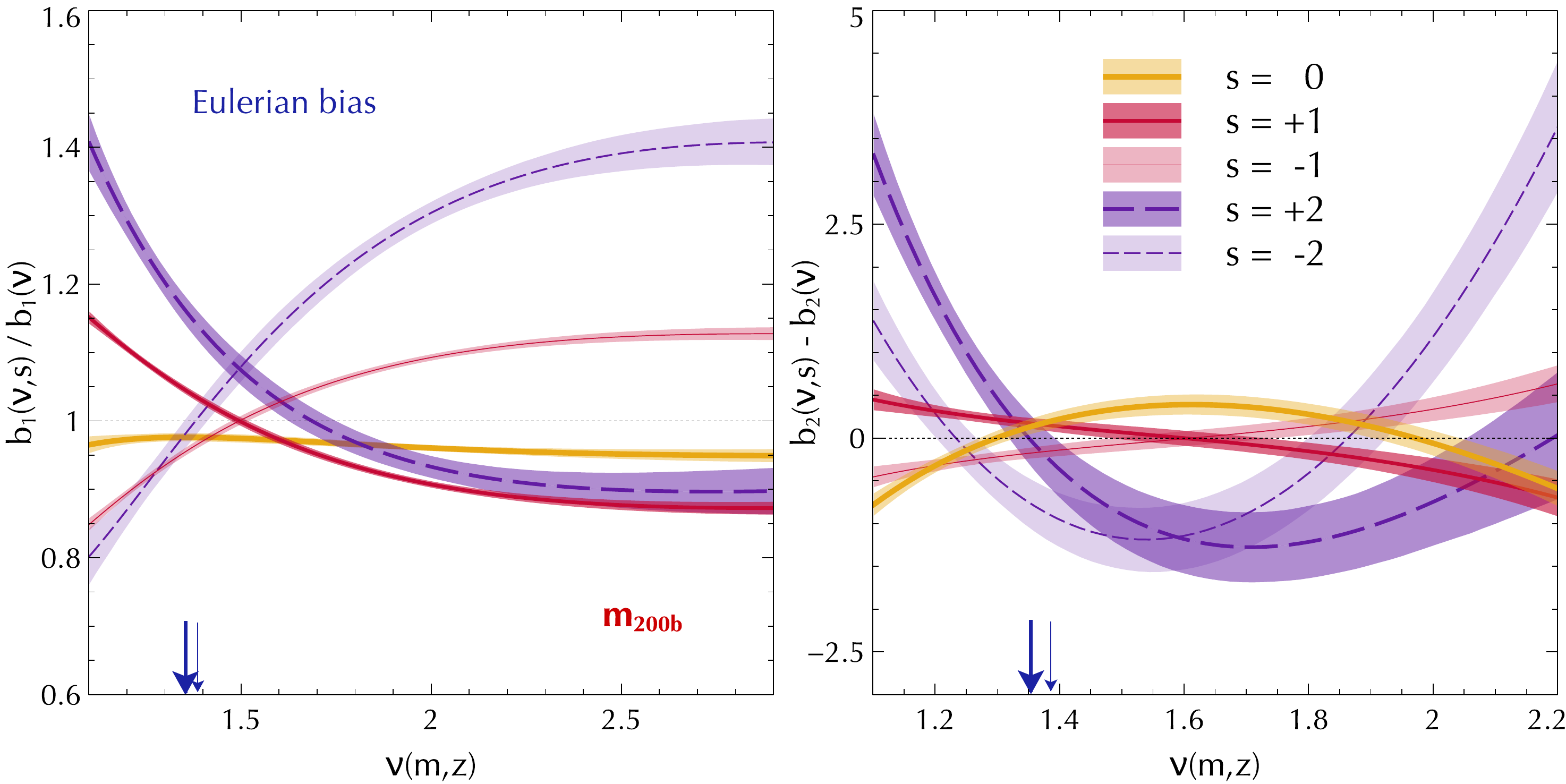}
\caption{Assembly bias at fixed standardized concentration $s$ (equation~\ref{eq:sdef}) from the Separate Universe technique assuming Lognormal concentrations (section~\ref{subsec:lognormal}). The curves show the ratio $b_1(\nu,s)/b_1(\nu)$ \emph{(left panel)} and the difference $b_2(\nu,s)-b_2(\nu)$ \emph{(right panel)} as functions of peak height $\nu$, obtained by using the polynomial fits to $\mu_n^{\rm L}$ and $\Sigma_n^{\rm L}$, $n=1,2$ (Figure~\ref{fig:abSUcoeffs} and Table~\ref{tab:ABcoeffs}), in \eqns{eq:b1Lag(m,s)}-\eqref{eq:b2Lag(m,s)} and \eqref{eq:localEulbias}. Results are shown for $s=0$ (thick solid yellow), $s=\pm 1$ (solid red) and $s=\pm 2$ (dashed purple). For each value of $|s| > 0$, thicker curves show results for positive $s$ and thinner curves for negative $s$. The results in the left panel used the universal T10 fitting form for the concentration independent linear bias $b_1(\nu)$. The bands represent $68.3\%$ error intervals obtained by Monte-Carlo sampling Gaussian distributions for the polynomial coefficients with the means and covariance matrices reported in Table~\ref{tab:ABcoeffs}. The arrows indicate values of $\nu_\ast(z)$ as in Figures~\ref{fig:abSUfitVsdirect} and~\ref{fig:abSUcoeffs}.}
\label{fig:abSU}
\end{figure*}

The errors on $\mu_n^{\rm L}$ and $\Sigma_n^{\rm L}$ are computed as follows. For each of our \NSIM\ realisations, we measure $\mu(m,z,\dL)$ and $\sigma^2(m,z,\dL)$ $50$ times using bootstrap resamplings of the $s$ values. The respective variances of these measurements are used as errors in fitting $4^{\rm th}$ order polynomials to these quantities. This gives us \NSIM\ estimates of $\mu_n^{\rm L}(m,z)$ and $\Sigma_n^{\rm L}(m,z)$; the mean of these is reported as the final measurement in Figure~\ref{fig:abSUcoeffs}, with error bars corresponding to the standard error on the mean over the \NSIM\ realisations.

Within the errors, all four linearly extrapolated coefficients are also consistent with being universal functions of $\nu(m,z)$. We therefore combine our measurements at the two redshifts and fit quadratic polynomials to $\mu_n^{\rm L}$ and $\Sigma_n^{\rm L}$, as functions of $y\equiv\log_{10}(\nu/1.5)$ in each case\footnote{We do not use the data points corresponding to the highest mass bin at $z=0$, since this contained fewer than 100 haloes per realisation in the fiducial simulation. Note also that, ignoring these data points, we see a clear oscillating trend in $\Sigma_1^{\rm L}$ (and also to a smaller extent in $\mu_1^{\rm L}$) that starts at the lowest masses. Milder versions of these oscillations are also seen in the all-halo bias measurements in Figures~\ref{fig:b1-SU} and~\ref{fig:b2-SU} (and were also present in the original L16 analysis; see, e.g., the green points in their Figure 1). These are very likely to be numerical artefacts of the measurement technique, and we therefore choose to wash them over using low order polynomial fits.}. We treat each of the four sets of coefficients independently, 
ignoring the covariances between, e.g., $\mu_1^{\rm L}$ and $\mu_2^{\rm L}$; this could in principle be improved by extracting the covariance structure at the level of the bootstrap analysis mentioned above.
The resulting best fit functions are shown as the smooth curves in Figure~\ref{fig:abSUcoeffs}, and the corresponding parameters are reported in Table~\ref{tab:ABcoeffs}.

These fitting functions for $\mu_n^{\rm L}$ and $\Sigma_n^{\rm L}$ with $n=1,2$ can be directly used in \eqns{eq:b1Lag(m,s)}-\eqref{eq:b2Lag(m,s)}, in conjunction with analytical fits/predictions for $b_1^{\rm L}(m,z)$ and $b_2^{\rm L}(m,z)$ to get smooth representations of assembly bias (with appropriate Lagrangian to Eulerian conversions where needed). The solid red curves in Figure~\ref{fig:abSUfitVsdirect} show the resulting linear Eulerian bias when averaged over $s>0.675$ and $s<-0.675$ weighted by the standard Gaussian distribution. We see that there is reasonable agreement between these Lognormal-based results and the results of direct binning from section~\ref{subsec:concbin}, with the inversion occurring at nearly the same value of $\nu$ in each case, although the Lognormal results are systematically lower than the directly binned ones. We discuss this further in section~\ref{sec:discuss}. Secondly, the errors on the Lognormal results, shown as the shaded bands (see below), are substantially smaller than the scatter in the directly binned results; this is consistent with the fact that (a) the former use the full halo population in any mass bin while the latter don't, and (b) the noise in the Lognormal results is  further reduced by using the polynomial fits described above.

More interestingly, we note that our analytical formulation above allows us to predict halo bias \emph{at fixed} $s$, without \emph{any} binning. We discuss the advantages of this formulation further in section~\ref{sec:discuss}. Figure~\ref{fig:abSU} shows the Eulerian ratio $b_1(m,z,s)/b_1(m,z)$ (left panel) and the Eulerian difference $b_2(m,z,s) - b_2(m,z)$ as functions of $\nu(m,z)$, for various fixed values of $s$. In addition to the polynomial fits for the assembly bias coefficients, we used the universal T10 fit for $b_1(\nu(m,z))$ in generating these curves. The error bands around these curves (as well as those in Figure~\ref{fig:abSUfitVsdirect}) were determined by Monte-Carlo sampling the Gaussian covariance matrix of the fit parameters of Table~\ref{tab:ABcoeffs} (and ignoring the errors on $b_1(\nu)$).

\begin{figure*}
\centering
\includegraphics[width=0.9\textwidth]{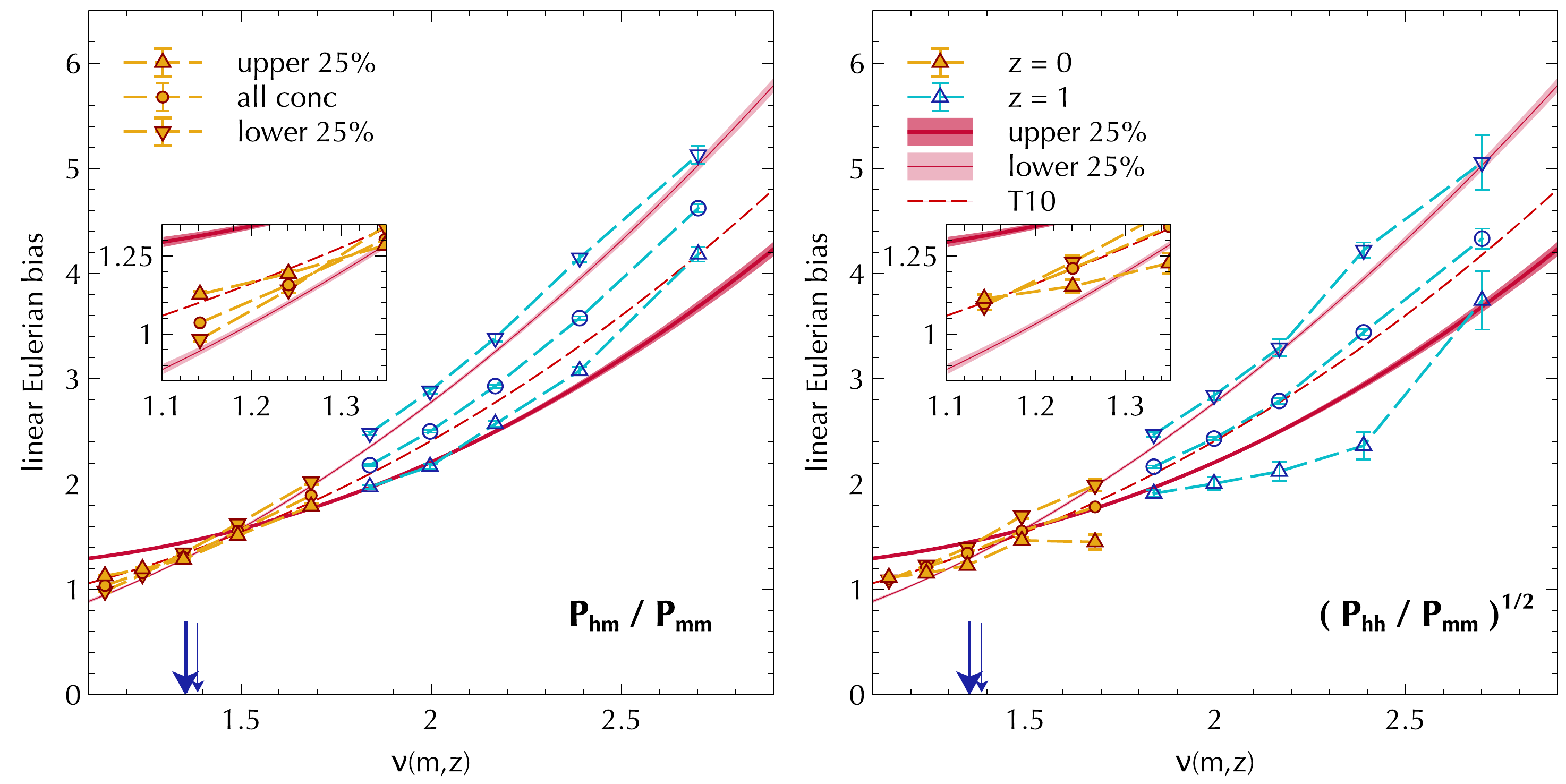}
\caption{Standard measurement of linear assembly bias using halo-matter \emph{(left panel)} and halo-halo power spectra \emph{(right panel)}. Filled yellow (open blue) symbols show results at $z=0$ ($z=1$), with upward and downward triangles respectively indicating measurements for haloes in the upper and lower quartiles of concentration, and circles indicating measurements for all haloes. For comparison, the Separate Universe measurements from Figure~\ref{fig:abSUfitVsdirect} are repeated in each panel. The arrows indicate the scales $\nu_\ast(z)$ as in previous Figures. The insets zoom into the low-mass behaviour of the $z=0$ measurements. See text for a discussion.}
\label{fig:abPk}
\end{figure*}

\subsection{Comparison with traditional estimates and previous results}
\label{subsec:traditional}
\noindent
For comparison, in Figure~\ref{fig:abPk} we show the linear halo bias determined from the auto- and cross-power spectra of haloes and dark matter in the $\dL=0$ simulation, split by halo mass and in three bins of concentration: all, upper quartile and lower quartile. We determined these bias values $b_1$ by fitting the function $b(k) = b_1+b_1^\prime k^2$ to the appropriate ratio of power spectra (indicated in the Figure labels) in the range $0.04 \lesssim k/(h\,{\rm Mpc}^{-1}) \lesssim 0.3$. 
We have checked that our high resolution box leads to statistically consistent results, indicating that these measurements are not affected by mass resolution.
The arrows again indicate the characteristic scales $\nu_\ast(z)$ obtained from the peak of the mass fraction (recall that the standard definition is $\nu_{\ast,{\rm std}}\equiv1$). 

We see that these standard measurements broadly agree with the Separate Universe results (repeated from Figure~\ref{fig:abSUfitVsdirect}), showing the same qualitative trends. In contrast with the latter, however, the inversion of the bias trend with halo concentration occurs at a scale $\nu=\nu_{\rm inv}$ such that $1<\nu_{\rm inv}\lesssim\nu_\ast$. Previous analyses that have studied the concentration dependence of assembly bias \citep{wechsler+06,jsm07,fw10} have typically reported values of $\nu_{\rm inv}\simeq1$. The difference between the inversion scales in the Separate Universe approach ($\nu_{\rm inv}\simeq1.5$) and, say, the results of \citet[][$m_{\rm inv}/m_{\ast,{\rm std}}\sim1.5$]{jsm07} is nearly an order of magnitude in mass. This difference is important to understand, since it could significantly alter, e.g., predictions for the impact of large scale environment on galaxy formation. We discuss possible reasons for these discrepancies in the inversion scale in section~\ref{sec:discuss}.

\section{Discussion}
\label{sec:discuss}
\noindent
The Separate Universe approach measures the response of small scale dynamics to changes in an infinite-wavelength overdensity. As such, our measurements of assembly bias using this approach represent the first genuinely long-wavelength correlations between assembly history and environment. Previous, traditional estimates of assembly bias have relied on ratios of correlation functions or power spectra at finite separations ranging from $\sim100\Mpch$ down to $\sim5\Mpch$. Any scale dependence of clustering -- particularly at scales $\lesssim10\Mpch$ \citep{shpl16} -- would consequently manifest as a mass dependence of the assembly bias signature. The measurements presented in this work are free from any scale dependence at scales smaller than our fiducial box size of $300\Mpch$. 

The Separate Universe approach also facilitates an analytical formalism (section~\ref{subsec:lognormal}) that  allows us to describe the concentration dependence of halo bias in a statistically self-consistent manner using convenient fitting functions \citep[see also][]{wechsler+06}. This makes our results directly applicable to analytical Halo Model studies of assembly bias. 
The dynamic reach in halo mass of the Separate Universe measurements is substantially larger than that of traditional estimators of halo bias in the same volume (Figures~\ref{fig:abSUcoeffs} and~\ref{fig:abPk}).
Below, we discuss some additional interesting aspects and potential applications of our results.

\subsection{Quadratic assembly bias}
\label{subsec:b2AB}
\noindent
Analytical treatments based on peaks theory \citep{dwbs08} and the excursion set approach \citep{cs13} naturally predict a correlation between halo environment and halo properties other than mass, such as peak curvature or halo concentration. As such, it is natural to expect that the effects of assembly bias should not be restricted to linear bias $b_1$. Traditional approaches using numerical simulations have been mostly limited to studying $b_1$, presumably due to the technical complication that extracting higher order coefficients $b_n$ from clustering measurements requires measuring $(n+1)$-point functions; splitting these further into percentiles of concentration, say, implies a drastic reduction in signal-to-noise\footnote{The only exception we are aware of is the study by \citet{abl08}, who used a cross-correlation technique based on counts-in-cells to extract the concentration dependence of local bias parameters $b_n$ with $n\leq4$. In particular, these authors reported a significant assembly bias signal in $b_2$ at $\nu\gtrsim2$, with \emph{no} signal for $1\lesssim\nu\lesssim2$. However, these results were based on a simulation with $m_{\rm part}\simeq5\times10^{11}\Mh$ and FoF haloes with $m\geq26m_{\rm part}$. Since this mass resolution is much coarser than the requirements for convergence of the concentration distribution (section~\ref{subsec:cmz_fid}), it is difficult to assess the significance of the discrepancies between the $\nu\simeq1$ results reported by \citet{abl08} and those presented in this work.}. On the other hand, given that standard analytical treatments \emph{do not} explain the inversion of the linear assembly bias trend around $m\sim m_\ast(z)$, it is of great interest to ask what simulations have to say regarding assembly bias at higher order.

The Separate Universe approach provides an ideal noise-reduction technique for accessing higher order bias coefficients. As our analysis above has demonstrated, measuring assembly bias effects at order $n$ is an almost trivial by-product of the method. And while the method is limited by the usual constraints of mass resolution and volume, these constraints are more relaxed than in traditional methods (Figures~\ref{fig:abSUcoeffs} and~\ref{fig:abPk}). 

\begin{figure}
\centering
\includegraphics[width=0.45\textwidth]{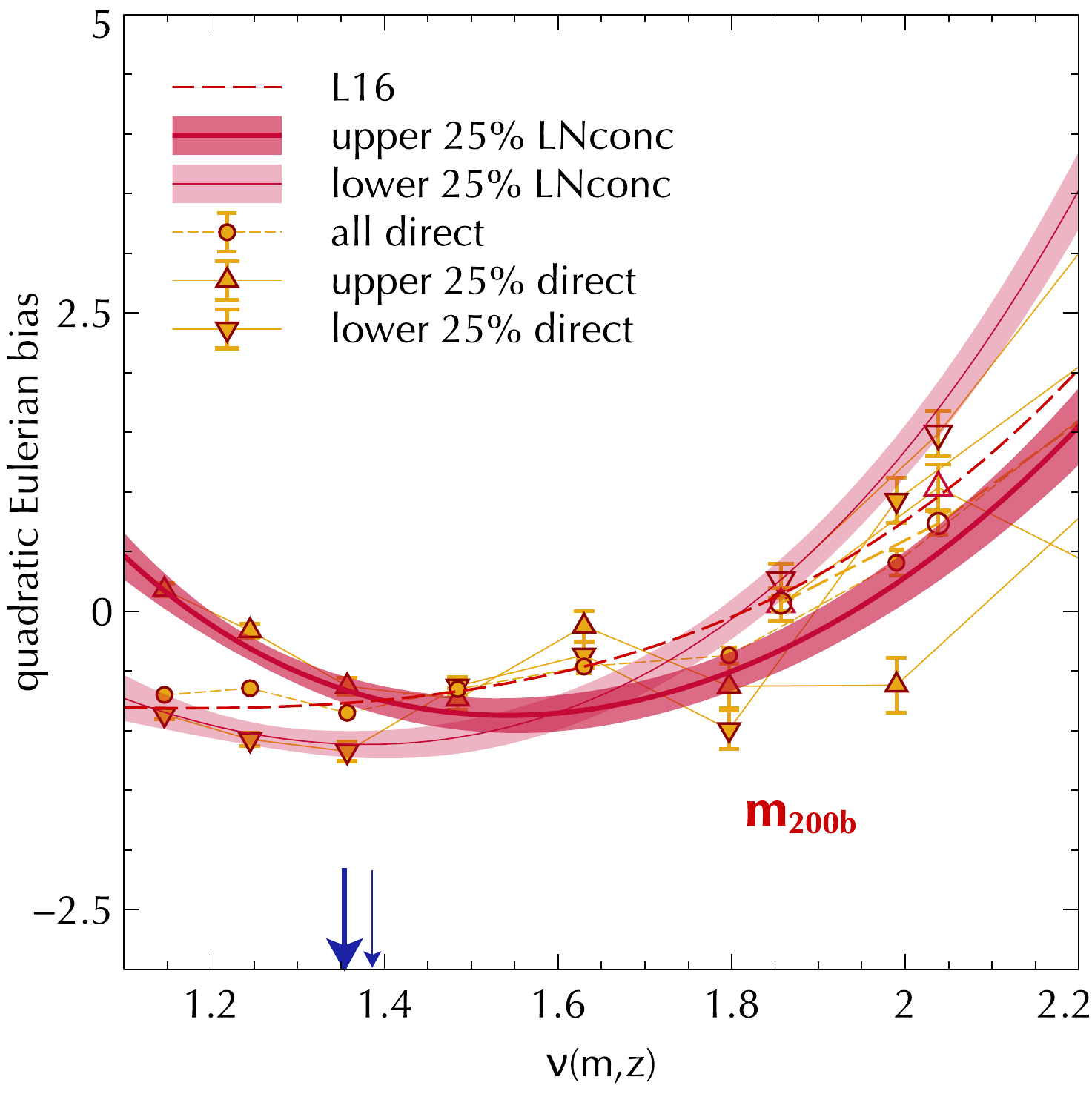}
\caption{Same as Figure~\ref{fig:abSUfitVsdirect}, for quadratic bias. The points with errors show the results of directly binning haloes in mass and concentration. For the Lognormal-based method, the dashed curve shows equation~5.2 of L16 for all haloes, while the thick (thin) solid curves show the result of averaging $b_2(m,s)$ (equations~\ref{eq:b1Lag(m,s)}-\ref{eq:b2Lag(m,s)} and~\ref{eq:localEulbias}) over $s > 0.675$ ($s < -0.675$).} 
\label{fig:abSUfitVsdirect-b2}
\end{figure}

The right panel of Figure~\ref{fig:abSU} shows our measurement of halo assembly bias at quadratic order, using the Lognormal assumption for halo concentrations. We see that the concentration dependence of $b_2$ has a distinctly richer and more nonlinear structure than that of $b_1$. This $s$-dependence is detected with high statistical significance, despite the high level of noise in measuring $\mu_2^{\rm L}$ and $\Sigma_2^{\rm L}$ (Table~\ref{tab:ABcoeffs}); this is mainly due to the substantial contribution of $\mu_1^{\rm L}$ and $\Sigma_1^{\rm L}$ -- which \emph{are} measured very accurately -- in \eqn{eq:b2Lag(m,s)}. The behaviour at large $\nu$ is qualitatively consistent with expectations from peaks theory, and there is also an inversion of the trend with $s$, as for $b_1$. This inversion, however, happens at somewhat larger $\nu$ for $b_2$ than for $b_1$. To test the robustness of these results, in Figure~\ref{fig:abSUfitVsdirect-b2} we compare them with measurements based on directly binning haloes in mass and standardized concentration (c.f. Figure~\ref{fig:abSUfitVsdirect}). The latter are considerably noisier, but consistent overall with the Lognormal results. It will be very interesting to compare these results with measurements of the concentration dependence of, e.g., the bispectrum between haloes and matter; we leave this exercise to future work.

\subsection{Assembly bias and halo mass}
\label{subsec:b1Vsb2}
\noindent
The strength of clustering of galaxies of different types (e.g., faint or bright, red or blue) is routinely used as an indicator of the typical masses of the respective host dark matter haloes, using HOD modelling \citep[see, e.g.,][]{zehavi+11}. For luminous galaxies, for example, this might involve comparing the large scale galaxy 2-point correlation function with the expected correlation function of dark matter haloes in some mass range; adjusting this mass range until the two correlation functions match, provides an estimate of the host halo mass. 

One concern regarding this methodology is that, if the galaxy sample selection has introduced a preference for high or low values of, say, host halo age or concentration, then the resulting assembly bias of the sample would introduce a bias in the inferred host properties \citep{zhv14}. For example, if a sample has preferentially picked low concentration haloes of mass $m\lesssim m_\ast$, any HOD analysis that ignores assembly bias could substantially underestimate the host halo mass, because the average bias is a weak function of mass in this regime. And since the host properties encoded by the HOD are (at least mildly) degenerate with the effects of changing cosmology \citep{rtwl14,more+13}, any such `assembly bias bias' would also potentially affect the accurate recovery of cosmological parameters from galaxy clustering.

\begin{figure}
\centering
\includegraphics[width=0.45\textwidth]{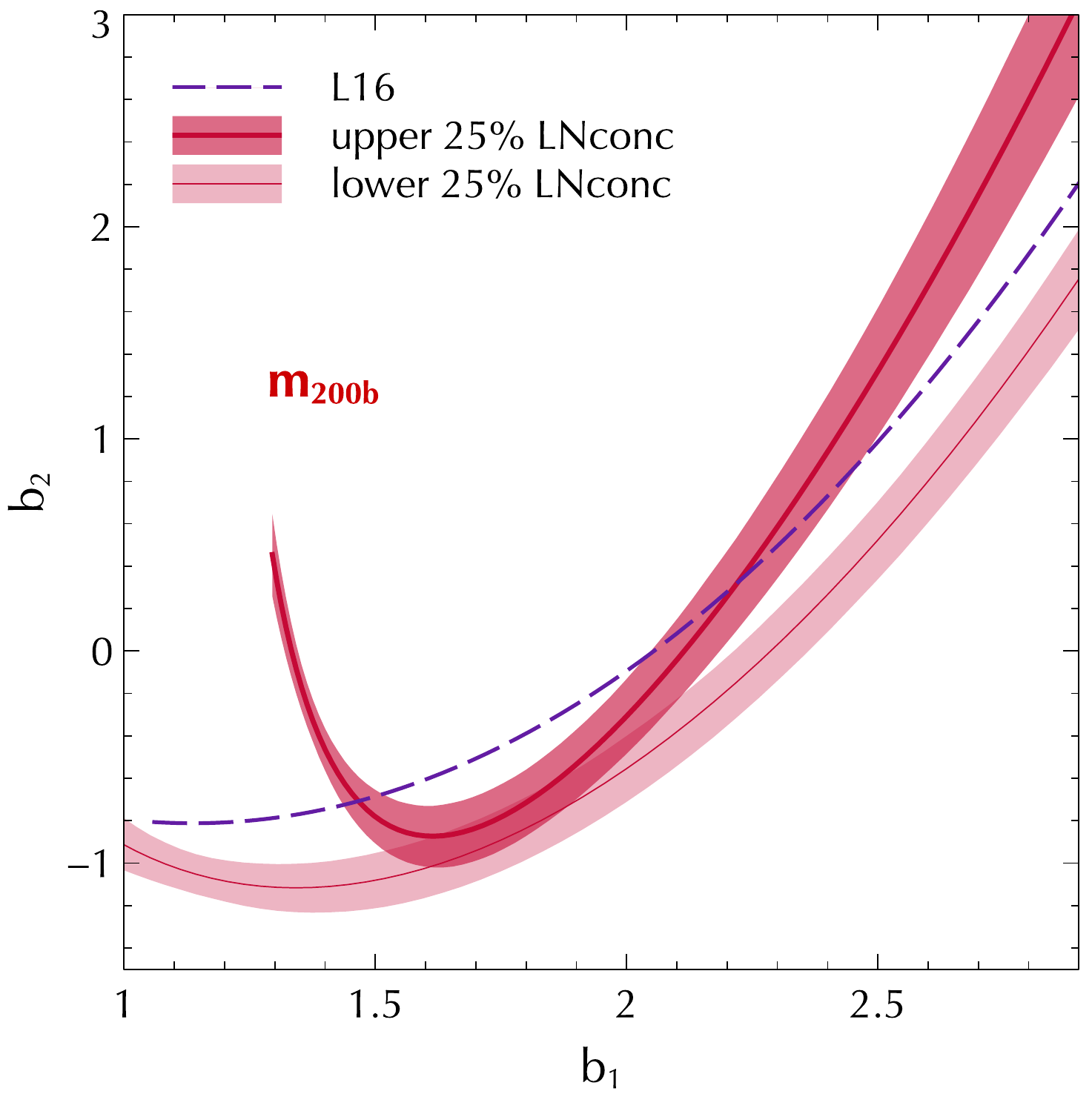}
\caption{Loci of $b_1$ against $b_2$ for all haloes (dashed curve, equation~5.2 of L16) and using our Lognormal-based results for haloes satisfying $s > 0.675$ and $s < -0.675$ (respectively, the thick and thin solid curves). Bands correspond to the $68.3\%$ errors on $b_2$ at fixed $b_1$, and we ignore these errors for the all-haloes curve. See text for a discussion.} 
\label{fig:abSU-b1Vsb2}
\end{figure}

In this context, assembly bias at quadratic order provides useful extra leverage in the problem, as we demonstrate next. Since the relation between $b_1$ and $b_2$ is known to be very close to universal when considering all haloes at a given $\nu$ \citep[][see also Figure~\ref{fig:b2-SU}]{hbg15,lwbs16}, the key idea here is to ask whether assembly bias introduces any \emph{non-universality} in this relation. To answer this, we use our Lognormal-based fits to plot $b_1$ against $b_2$ for haloes selected to be in the upper or lower quartiles of standardized concentration $s$. The results are shown, respectively, as the thick and thin solid lines with error bands in Figure~\ref{fig:abSU-b1Vsb2}. These are clearly different from each other and also from the average relation calibrated by L16 and shown as the dashed line. For $b_1\lesssim1.5$, in particular, the behaviour of $b_2$ is a strong function of concentration. 

This suggests that simultaneous measurements of $b_1$ and $b_2$ for a given galaxy sample can be very useful in (a) \emph{detecting} assembly bias and (b) \emph{mitigating} any effects of assembly bias in estimating halo mass. For example, comparing such measurements with calibration plots such as Figure~\ref{fig:abSU-b1Vsb2} would allow for an estimate of the typical value of $s$ of the host halo population, which could then be combined with the measured value of $b_1$, say, to read off halo mass (or $\nu$) from Figure~\ref{fig:abSUfitVsdirect}. And while we have chosen to display results in quartiles of $s$, our analytical formalism allows us to calibrate assembly bias at \emph{fixed} $s$ (Figure~\ref{fig:abSU} and equations~\ref{eq:b1Lag(m,s)}-\ref{eq:b2Lag(m,s)}). The only limitation is the error on the calibration; apart from the systematic uncertainties discussed below, this can be easily reduced using simulations with larger volume and/or higher mass resolution.

\subsection{Assembly bias and tides}
\label{subsec:tides}
\noindent
The observation that old, highly concentrated haloes of low mass cluster more strongly than younger, less concentrated haloes of the same mass is a challenge to analytical models based on peaks theory and/or the excursion set approach. A compelling argument for explaining the physical reason behind this trend was put forth by \citet{hahn+09}, who showed that the assembly bias of low mass haloes was consistent with being driven by the \emph{tidal} influence of nearby high-mass objects. This influence can be in the form of redirected mass flows (e.g., along filaments) which prevent small haloes from accreting mass, and also more directly in the form of tidal stripping of mass from a small halo that passes through and eventually exits a larger one. \citet{hahn+09} demonstrated that both mechanisms occur in simulations, and since nonlinear gravitational evolution produces a strong correlation between small-scale tides and large scale density, these mechanisms would then correlate with the latter and produce assembly bias \citep[see also][]{bprg16}. 

These ideas are consistent with our Separate Universe measurements of assembly bias. To see this, consider that this mechanism of tidally truncated mass assembly (or the more extreme tidal stripping of `splashback' haloes) amounts to saying that old, small objects are old and small because they were affected by nearby massive objects. The \emph{abundances} of old, small objects are therefore positively correlated with those of massive objects. A simple back-of-the-envelope calculation then shows that the large scale \emph{bias} (i.e., response of abundances to large scale density) of old, small objects will be enhanced compared to the mean bias of small objects, by an amount determined by the bias of the \emph{large} objects that tidally influenced the smaller ones. It is this response that is being measured by the Separate Universe analysis.

One could go even further, and use the Separate Universe technique to probe the influence of tides directly. There is already considerable interest in extending this technique to measure the response of halo counts to \emph{large scale} tides -- this would measure the so-called `non-local' bias that has been recently detected using traditional techniques \citep{chan/scoccimarro/sheth:2012,baldauf/seljak/etal:2012,Saito+14}. Assuming that the technical hurdles of self-consistently simulating infinite-scale anisotropic tides can be overcome \citep{is16}, the same technique could be extended to smaller boxes to directly study, under controlled circumstances, the tidal mechanisms discussed by \citet{hahn+09}. As an aside, we note that the analytical formalism developed in this work can be easily expanded to include response coefficients to tidal effects, organised for example by powers of the rotational invariants of the tidal tensor \citep{gpp12,scs13,desjacques13}. We will explore these ideas in future work.

\subsection{Quantitative aspects of the assembly bias signature}
\label{subsec:systematics}
\noindent
Our combination of the Separate Universe technique and an analytical formalism based on the Lognormal assumption of halo concentrations has allowed us to achieve unprecedented precision in the calibration of assembly bias. There are, however, some aspects of our measurements that deserve further discussion. We briefly address these issues here, leaving a fuller exploration to future work.

\subsubsection{Binned concentrations versus the Lognormal assumption}
\label{subsubsec:binVsLN}
\noindent
The comparison between the Lognormal-based results and those of direct binning in Figures~\ref{fig:abSUfitVsdirect} show that, although both methods show an inversion in the assembly bias trend for $b_1$ at $\nu=\nu_{\rm inv}\simeq1.5$, the results of direct binning tend to be systematically below the Lognormal ones. This is most likely caused by the distribution of $s$ being only approximately Gaussian (Figure~\ref{fig:s_hist}). In principle, this non-Gaussianity could also be calibrated, with the resulting coefficients $\Cal{P}^{\rm L}_n(m,s)$ (equation~\ref{eq:Pn-avg}) then being used in \eqn{eq:<bn|m,s>}. This calibration could be done using, e.g., Gaussian mixtures, or more simply through a `Gaussianization' by finding a function $G$ such that $s_G = G(s)$ is Gaussian distributed. The latter idea is attractive since our analytical formalism would then be exact when expressed in terms of $s_G$ and, moreover, the same technique could be extended, in principle, to \emph{any} assembly bias variable.

A more interesting question is what causes this non-Gaussianity in $p(s|m,\dL)$. We have checked that the shapes of these distributions are essentially independent of mass definition, and also relatively insensitive to the virial ratio cut discussed in section~\ref{subsec:SUtech}. It would be very interesting to identify one or more halo properties that define the small fraction of haloes in the negative $s$ tail of each $p(s|m,\dL)$ seen in Figure~\ref{fig:s_hist}. For the time being, we conclude that the small difference between the binned and Lognormal results in Figure~\ref{fig:abSUfitVsdirect} reflects the level of systematic uncertainty inherent in our procedure.

\subsubsection{Assembly bias inversion scale}
\label{subsubsec:invscale}
\noindent
As mentioned previously, our Separate Universe measurements of $b_1(\nu,s)$ show an inversion of the assembly bias trend around $\nu_{\rm inv}\simeq1.5$, which is substantially higher than the inversion scale $\nu_{\rm inv}\simeq1$ seen by most previous studies (at $z=0$, for example, this corresponds to nearly an order of magnitude difference in halo mass). Here we explore possible reasons for this discrepancy.

At the outset, we note that the scale $\nu=1$ is special only in that the original Press-Schechter mass fraction $\sim\nu\,{\rm e}^{-\nu^2/2}$ has its maximum at this scale \citep{ps74}. The mass fraction in simulations actually peaks at substantially higher masses, as we have already seen in Figure~\ref{fig:mf}, with a characteristic mass $m_\ast(z)$ that corresponds to $\nu_\ast(z)\sim1.35$ (see, e.g., Figure~\ref{fig:abSUfitVsdirect}). If the inversion of assembly bias is indeed physically tied to the characteristic mass, then it would be more natural to expect $\nu_{\rm inv}\sim\nu_\ast$ -- as we do, in fact, see in the Separate Universe results ($\nu_{\rm inv}\simeq1.5$) as well as in the traditional estimate based on the halo-matter cross-power spectrum ($\nu_{\rm inv}\simeq1.3$, left panel of Figure~\ref{fig:abPk}). In the absence of a robust analytical model for assembly bias, however, it is difficult to make more precise statements. We also note that the inversion scale appears to be a strong function of the variable being used to study assembly bias \citep[see, e.g.,][]{jsm07,fw10}.

To try and understand why our results differ from those of previous authors who also used halo concentration as the assembly bias variable, it is worth listing the differences between those analyses and ours. Firstly, we have used haloes identified using \textsc{rockstar} with masses given by the Spherical Overdensity (SO) definition, while previous analyses have typically used haloes identified by, and masses assigned by, the standard FoF algorithm \citep{jsm07,fw10}. The clustering of FoF and SO haloes is known to be different at the $\sim20\%$ level at the same mass \citep{Tinker10}, so it would not be surprising if the assembly bias trends for these mass definitions also differ in quantitative detail. The study by \citet{wechsler+06}, on the other hand, did use SO haloes (with $m_{\rm vir}$ rather than $m_{\rm 200b}$ masses) and also found an inversion at $\nu_{\rm inv}\simeq 1$, although the noise in that measurement would allow for uncertainty in the inversion mass scale of factors of a few. Overall, then, halo mass definition may play some role in explaining the differences in inversion scale.

Turning to the definition of assembly bias itself, we note that previous analyses have typically fit constants to ratios of correlation functions at rather small scales, with separations $5\lesssim r / (\Mpch) \lesssim 20$  \citep{wechsler+06,fw10}. This is also true of our own comparison analysis in section~\ref{subsec:traditional}, which used scales $\sim10$-$85\Mpch$ in Fourier space. One notable exception is the study by \citet{jsm07}, who used high resolution simulations with very large volume and were therefore able to probe scales $\sim35$-$200\Mpch$ in Fourier space. The limitations of sample variance mean that the signal-to-noise in all these measurements is dominated by the smallest scales being probed. Consequently, one cannot rule out the possibility that \emph{all} of these analyses were affected by at least a small  level of scale dependence of halo bias; fitting constants to scale dependent measurements would then convert this scale dependence into a \emph{mass} dependence of the assembly bias signature. 
This is also evident in the difference between our own results using auto- and cross-power spectra for the same haloes in Figure~\ref{fig:abPk}, with $\nu_{\rm inv}$ in the former case ($\simeq1.1$) being smaller than in the latter ($\simeq1.3$). Since our results used a fit to a constant plus a term proportional to $k^2$, and should therefore be less susceptible to scale dependence than those based on fitting a constant alone, this difference between the auto- and cross-power results is likely to represent a lower bound on systematic effects in traditional estimates of assembly bias.
Making a more accurate assessment of this possible contamination due to scale dependence is, however, beyond the scope of the present work.

\section{Summary}
\label{sec:conclude}
\noindent
We have presented a high precision calibration of halo assembly bias -- specifically, the dependence of halo concentration on large scale environment -- using the Separate Universe technique. Our main results are the following:

\begin{itemize}
\item We have developed an analytical formalism (section~\ref{subsec:lognormal}) that exploits the near-Lognormal distribution of halo concentrations at fixed mass and allows for an extremely precise measurement (Figure~\ref{fig:abSU}) of the dependence of halo bias on standardized concentration $s$ (equation~\ref{eq:sdef}). 
\item When averaged over percentiles of $s$, the results based on our analytical formalism are in good agreement with those based on directly binning haloes in percentiles of concentration at fixed mass (Figures~\ref{fig:abSUfitVsdirect} and~\ref{fig:abSUfitVsdirect-b2}). Within errors, we find that the mass and redshift dependence of assembly bias is universal, being a function of peak height $\nu(m,z)$ (Figure~\ref{fig:abSUcoeffs}). 
\item The combination of the Separate Universe technique and our analytical formalism allows for an almost trivial extension to high order bias parameters $b_n$, and we have presented calibrations of assembly bias at linear and  quadratic order, $b_1(\nu,s)$ and $b_2(\nu,s)$ (equations~\ref{eq:b1Lag(m,s)}-\ref{eq:b2Lag(m,s)}; Table~\ref{tab:ABcoeffs}; Figures~\ref{fig:abSUcoeffs} and~\ref{fig:abSU}).
\item While the locus of $b_1$ against $b_2$ for all haloes is independent of redshift \citep[][see also Figure~\ref{fig:b2-SU}]{lwbs16}, we find that it \emph{does} depend strongly on halo concentration, particularly at small $b_1$ (Figure~\ref{fig:abSU-b1Vsb2}). This suggests that assembly bias could be observationally detected, and its effects on cosmological analyses be mitigated, by simultaneously measuring $b_1$ and $b_2$ for a galaxy sample (section~\ref{subsec:b1Vsb2}).
\item A comparison of our results for $b_1$ with traditional estimates based on ratios of power spectra shows that the inversion of the linear assembly bias trend with concentration occurs at substantially higher mass scales in the Separate Universe approach than in the traditional approaches (Figure~\ref{fig:abPk}). Apart from differences due to halo mass definition, this could be due to residual scale dependence of halo bias in the traditional approaches; this is explicitly absent in the Separate Universe technique which measures the response of halo abundances to an infinite-wavelength overdensity (section~\ref{subsubsec:invscale}).
\end{itemize}

We note that a recent preprint \citep{lms17}, which appeared soon after ours, also explores assembly bias with the Separate Universe technique, using several variables including halo concentration. Their results, where they overlap with our work, are consistent with our findings above.

In future work, we will return to several aspects of our analysis that deserve further exploration. These include, a better calibration of the non-Gaussian shape of $p(s|m,\dL)$, an extension of the formalism and simulations to include tidal effects, and a validation of the results for assembly bias in $b_2$ using measurements of the bispectrum between haloes and matter as a function of halo concentration.

\section*{Acknowledgements}
\noindent
We thank Oliver Hahn for many useful discussions. We gratefully acknowledge the use of high performance computing facilities at IUCAA, Pune.
The plots in this work were generated using the software \textsc{Veusz} (http://home.gna.org/veusz/).
The research of AP is supported by the Associateship Scheme of ICTP, Trieste and the Ramanujan Fellowship awarded by the Department of Science and Technology, Government of India.

\bibliography{masterRef}

\appendix

\section{Initial Conditions}
\label{app:ICs}
\noindent
In principle, simulating a cosmology with non-zero spatial curvature means that the Fourier basis is not strictly valid, and one must work with eigenfunctions of the Helmholtz operator $\nabla^2 - \Cal{K}$ instead, where $\Cal{K}=-\Omega_{\rm k}/H_0^2$. However, for the range of \dL\ values used in this work, the corresponding curvature scales are large enough compared to our chosen box size that we do not need to worry about this complication. Since the comoving densities of dark matter and baryons, represented by the combinations $\Omega_{\rm cdm}h^2$ and $\Omega_{\rm b}h^2$, respectively, are also held constant in the calculation, the shape of the initial transfer function in all Separate Universe cosmologies is identical to the one in the fiducial cosmology \citep[][hereafter, W15]{wsck15a}. The only difference is in the normalisation of the power spectrum, which must be adjusted to account for the different linear growth histories across cosmologies, which lead to different values of the growth factor $D(a(t))$ at the same cosmic time $t$. We discuss this below, essentially following the description in W15 with a modification due to our use of a normalisation based on $\sigma_8$. 

The dimensionless matter power spectrum $\Delta^2(k,t)\equiv k^3P(k,t)/2\pi^2$ at comoving wavenumber $k$ and scale factor $a(t)$ in the post-radiation era (assuming decaying modes have died away) can be written as
\begin{align}
\Delta^2(k,t) &= D^2(a(t)) \Cal{A}_{\rm s} \left(\frac{k}{k_0}\right)^{n_{\rm s}-1}\,\left(\frac{k^2}{\Omega_{\rm m}H_0^2}\right)^2\,T^2(k)
\notag\\
&\equiv D^2(a(t))\,\Cal{T}^2(k)
\label{eq:app:Delta2k}
\end{align}
where $D(a)$ is the linear theory growth factor, $\Cal{A}_{\rm s}$ is the amplitude of scalar perturbations, $T(k)$ is the transfer function, $k_0$ is a constant pivot scale \citep{Komatsu2010} and the other parameters have their usual meanings as described in Section~\ref{subsec:SUtech}. 
The second line isolates the time-independent part of the power spectrum into the quantity $\Cal{T}^2(k)$.

We wish to modify the fiducial power spectrum to the one in the Separate Universe defined by a non-zero \dL; following W15, we will denote all quantities in the new cosmology using a tilde. Then we have $\tilde{\Cal{A}}_{\rm s} = \Cal{A}_{\rm s}$, $\tilde\Omega_{\rm m}\tilde h^2 = \Omega_{\rm m}h^2$ and $\tilde T(k) = T(k)$, so that $\tilde{\Cal{T}}^2(k) = \Cal{T}^2(k)$, but $\tilde D(\tilde a(t))\neq D(a(t))$ for a given cosmic time $t$. Denoting $\tilde a_{\rm in} = \tilde a(t_{\rm in})$ and $a_{\rm in}=a(t_{\rm in})$, the new power spectrum at the initial time $t_{\rm in}$ then satisfies
\be
\tilde\Delta^2(k,t_{\rm in}) = \tilde D^2(\tilde a_{\rm in}) \Cal{T}^2(k) =  \left(\frac{\tilde D(\tilde a_{\rm in})}{D(a_{\rm in})}\right)^2 \times \Delta^2(k,t_{\rm in})
\label{eq:app:Deltatilde-init}
\ee
For the fiducial cosmology, we use \textsc{camb} to generate the power spectrum at $a(t)=1$. By default, this is normalised by \textsc{music} using a specified value of $\sigma_8$. In practice, this means that \textsc{music} sets the $a=1$ power spectrum to
\be
\Delta^2(k,a=1) = \frac{\sigma_{8\Mpch}^2\Cal{T}^2(k)}{\int\der\ln k\,\Cal{T}^2(k)\,W_{\rm TH}^2(k\cdot8\Mpch)} \,,
\label{eq:app:sig8fid-1}
\ee
where $W_{\rm TH}(kR)$ is the Fourier transform of a spherical tophat filter of comoving radius $R$, and we have been careful to spell out the meaning of $\sigma_8$. The normalised power spectrum is then extrapolated backwards to the initial time by \textsc{music} using the fiducial growth function $D(a)$: 
\be
\Delta^2(k,t_{\rm in}) = \left(\frac{D(a_{\rm in})}{D(a=1)}\right)^2\,\Delta^2(k,a=1)\,.
\label{eq:app:musicnorm-fid}
\ee
When working with the Separate Universe cosmology, \textsc{music} uses the input $\tilde\sigma_8$ \emph{assuming that it is defined at} $\tilde a=1$, which is not the same cosmic time at which the fiducial $a=1$. In addition, it uses the Hubble constant $\tilde h$ of the new cosmology in performing the integral weighted by the tophat filter, and finally extrapolates to the specified initial time using the growth factor of the new cosmology. Keeping all this in mind, let us define the input $\tilde\sigma_8$ of the new cosmology using
\begin{align}
\tilde{\sigma}_{8\tilde{h}^{-1}{\rm Mpc}}^2 &= \sigma_{8\Mpch}^2 \left(\frac{\tilde{D}(\tilde{a}=1)}{D(a=1)}\right)^2\notag\\
&\ph{\sigma_{8Mpc}^2}\times
\frac{\int\der\ln k\,\Cal{T}^2(k)\,W_{\rm TH}^2(k\cdot8\tilde {h}^{-1}{\rm Mpc})}{\int\der\ln k\,\Cal{T}^2(k)\,W_{\rm TH}^2(k\cdot8\Mpch)}\,.
\label{eq:app:sig8tilde}
\end{align}
This guarantees that the $\tilde a=1$ power spectrum as computed by \textsc{music} is given by
\begin{align}
\tilde\Delta^2&(k,\tilde a=1) \notag\\
&= \frac{\tilde\sigma_{8\tilde{h}^{-1}{\rm Mpc}}^2\Cal{T}^2(k)}{\int\der\ln k\,\Cal{T}^2(k)\,W_{\rm TH}^2(k\cdot8\tilde{h}^{-1}{\rm Mpc})} \notag\\
&= \left(\frac{\tilde{D}(\tilde{a}=1)}{D(a=1)}\right)^2\,\frac{\sigma_{8\Mpch}^2\Cal{T}^2(k)}{\int\der\ln k\,\Cal{T}^2(k)\,W_{\rm TH}^2(k\cdot8\Mpch)}\notag\\
&= \left(\frac{\tilde{D}(\tilde{a}=1)}{D(a=1)}\right)^2\,\Delta^2(k,a=1)\,,
\label{eq:app:Deltatilde-Delta}
\end{align}
which in turn leads \textsc{music} to compute the initial power spectrum of the new cosmology as
\begin{align}
\tilde\Delta^2(k,t_{\rm in}) &= \left(\frac{\tilde D(\tilde a_{\rm in})}{\tilde D(\tilde a=1)}\right)^2\,\tilde \Delta^2(k,\tilde a=1)\notag\\
&= \left(\frac{\tilde D(\tilde a_{\rm in})}{D(a=1)}\right)^2\,\Delta^2(k,a=1)\notag\\
&= \left(\frac{\tilde D(\tilde a_{\rm in})}{D(a_{\rm in})}\right)^2\,\Delta^2(k,t_{\rm in})\,,
\label{eq:app:Delta-final}
\end{align}
exactly as required by \eqn{eq:app:Deltatilde-init}. Note that this normalisation is accomplished by the single \eqn{eq:app:sig8tilde} which calculates $\sigma_8$ in the new cosmology; all the other calculations are performed internally by \textsc{music} without any input from the user. This completes the discussion on setting initial conditions in the Separate Universe cosmologies.

\section{Correcting for NFW fitting errors}
\label{app:NFWfitcorr}
\noindent
We assume that statistical errors incurred in fitting NFW profiles to haloes add in quadrature to the intrinsic scatter of halo concentrations. The constant, corrected values for intrinsic scatter reported by previous authors should then be a lower bound on our measured values of $\sigma_0^2(m,z)$ in the fiducial simulation (see equation~\ref{eq:cbarsig0def}). The right panel of Figure~\ref{fig:cmz} shows that the value reported by \citet{wechsler+02} is, in fact, consistent with such a floor\footnote{The fact that the \citet{bullock+01} value is substantially higher was explained by \citet{wechsler+02} as being an oversight in the former paper which erroneously reported the uncorrected value of the scatter. It is not clear to us, however, why the value reported by \citet{dk15} is larger than our measurements at large masses.}. 

In the following, we will assume that the intrinsic scatter $\sigma^2_{\rm lnc}$ in the fiducial simulation is \emph{constant} with a value given by \citet{wechsler+02}, and that \emph{all} the mass and redshift dependence that we measure in $\sigma_0^2$ is due to fitting errors. This allows us to statistically correct our measurements of the assembly bias coefficients $\mu_n^{\rm L}$ and $\Sigma_n^{\rm L}$ as described below. For notational ease, we will suppress the mass and redshift dependence throughout this section. We also assume that the intrinsic distribution of the log-concentrations is exactly Gaussian, as is the distribution of the fitting noise.

Let us write the log-concentration of the fiducial cosmology as
\be
\ln c = \overline{\ln c} + \sigma_{\rm lnc}\left(s + \varepsilon\right)\,,
\label{eq:app:lnc_fid}
\ee
where $s$ is the true, standardized log-concentration (zero mean and unit variance) and $\varepsilon$ is the contribution of the fitting noise (assumed to be Gaussian) with zero mean and standard deviation $\sigma_\varepsilon$. The measured mean is therefore unbiased, while the measured variance picks up a contribution due to $\sigma^2_\varepsilon$:
\be
\avg{\ln c|\dL=0} = \overline{\ln c}\,;\quad {\rm Var}(\ln c|\dL=0) = \sigma^2_{\rm lnc}\left(1 + \sigma^2_\varepsilon\right)\,.
\label{eq:app:meanvarlnc_fid}
\ee
Assuming that the fitting noise $\varepsilon$ has a distribution that depends only on mass and redshift \emph{but not on} \dL, we can now write similar results for a Separate Universe simulation with $\dL\neq0$. The log-concentration in such a simulation satisfies
\be
\ln c = \overline{\ln c}(\dL) + \sigma_{\rm lnc}(\dL)\,s + \sigma_{\rm lnc}\,\varepsilon\,,
\label{eq:app:lnc_su}
\ee
and our estimator for the standardized variable becomes
\begin{align}
\hat s &\equiv \frac{\ln c - \avg{\ln c|\dL=0}}{\sqrt{{\rm Var}(\ln c|\dL=0)}} \notag\\
&= \frac{\left\{\left(\overline{\ln c}(\dL)-\overline{\ln c}\right)/\sigma_{\rm lnc}\right\} + s\,\left\{\sigma_{\rm lnc}(\dL)/\sigma_{\rm lnc}\right\} + \varepsilon }{\sqrt{1 + \sigma^2_\varepsilon}}
\label{eq:app:s_est_su}
\end{align}
This gives us
\be
\avg{\hat s} = \frac{\left(\overline{\ln c}(\dL)-\overline{\ln c}\right)}{\sigma_{\rm lnc}\sqrt{1 + \sigma^2_\varepsilon}} \equiv \mu(\dL)\,,
\label{eq:app:<s|dL>}
\ee
and
\be
{\rm Var}(\hat s) = \frac{\sigma^2_{\rm lnc}(\dL)/\sigma^2_{\rm lnc} + \sigma^2_\varepsilon}{1 + \sigma^2_\varepsilon} \equiv \sigma^2(\dL)\,.
\label{eq:app:Var(s|dL)}
\ee
The `true' values of this mean and variance would be given by the same expressions upon setting $\sigma_\varepsilon\to0$. Moreover, these biased estimators nevertheless satisfy $\mu(\dL=0)=0$ and $\sigma^2(\dL=0)=1$ by construction, exactly like their unbiased counterparts. It is then easy to see that the \emph{Taylor coefficients} of the measured (i.e., biased) and true (unbiased) functions, defined in \eqn{eq:munSigndef}, satisfy
\begin{align}
\mu_n^{\rm (unbiased)} &= \mu_n^{\rm (measured)}\times\sqrt{1+\sigma^2_\varepsilon} \notag\\
\Sigma_n^{\rm (unbiased)} &= \Sigma_n^{\rm (measured)}\times\left(1+\sigma^2_\varepsilon\right)\,,
\label{eq:app:Taylor_corr}
\end{align}
for $n\geq1$. In practice, we use polynomial fits to the scatter of the log-concentrations in the fiducial cosmology as described in Section~\ref{subsec:cmz_fid} to estimate $\sigma^2_\varepsilon(m,z)$, which we then use in \eqn{eq:app:Taylor_corr} together with (biased) measurements of $\mu_n^{\rm L}(m,z)$ and $\Sigma_n^{\rm L}(m,z)$ to obtain unbiased estimates. 

This is equivalent to directly correcting the values of $\hat s$ by a multiplicative factor $\sqrt{1+\sigma^2_\varepsilon}$ prior to estimating the Taylor coefficients of its mean and variance, with the only difference that the $\dL\to0$ limit of the variance would now be $1+\sigma^2_\varepsilon$ instead of unity. We used this second method for correcting the concentrations when reporting the results of Separate Universe assembly bias from direct binning (section~\ref{subsec:concbin} and Figure~\ref{fig:abSUfitVsdirect}).

\section{The Hermite Expansion}
\label{app:LNcalc}
\noindent
The strategy to derive \eqns{eq:b1Lag(m,s)} and~\eqref{eq:b2Lag(m,s)} is to set $p(s|m,\dL)$ in \eqn{eq:p(s|d)taylor} to be a Gaussian in $s$ with mean $\mu(m,\dL)$ and variance $\sigma^2(m,\dL)$, replace these in terms of their Taylor expansions from \eqn{eq:munSigndef} and then expand the l.h.s. of \eqn{eq:p(s|d)taylor} in powers of \dL, thereby allowing us to read off the expressions for $\Cal{P}_n^{\rm L}(m,s)$. We sketch the details below, for $n=1,2$.

For clarity, in the following we will suppress the \dL- and/or mass-dependence of quantities, whenever no confusion can arise. The Gaussian assumption for $p(s|m,\dL)$ gives us
\begin{align}
\frac{p(s|m,\dL)}{p(s|m,\dL=0)}
&= \frac{{\rm e}^{-(s-\mu)^2/2\sigma^2}/\sqrt{2\pi\sigma^2}}{{\rm e}^{-s^2/2}/\sqrt{2\pi}}\notag\\
&= \frac{{\rm e}^{-\mu^2/2\sigma^2}}{\sqrt{\sigma^2}}\,{\rm e}^{s\mu/\sigma^2}\,{\rm e}^{-s^2\left(1/\sigma^2-1\right)/2}\,,
\label{eq:gaussratio}
\end{align}
To simplify the calculation, we can first Taylor expand the combinations $\mu/\sigma^2$, $\mu^2/\sigma^2$, $1/\sigma^2$ and $1/\sqrt{\sigma^2}$ in powers of \dL, using the expressions in \eqn{eq:munSigndef}. Keeping terms up to order $\dL^2$, this gives us
\begin{align}
\mu/\sigma^2 &= \mu_1\dL + \left(\mu_2-2\mu_1\Sigma_1\right)\dL^2/2 + \ldots \\
\mu^2/\sigma^2 &= \mu_1^2\dL^2 + \ldots \\
1/\sigma^2 &= 1 - \Sigma_1\dL + \left(2\Sigma_1^2-\Sigma_2\right)\dL^2/2 + \ldots \\
1/\sqrt{\sigma^2} &= 1 - \Sigma_1\dL/2 + \left(3\Sigma_1^2/2-\Sigma_2\right)\dL^2/4 + \ldots
\end{align}
The exponential factors in \eqn{eq:gaussratio} then become
\begin{align}
\frac{{\rm e}^{-\mu^2/2\sigma^2}}{\sqrt{\sigma^2}} &= 1 - \frac{\Sigma_1}{2}\dL + \left(\frac{3\Sigma_1^2}{4}-\frac{\Sigma_2}{2}-\mu_1^2\right)\frac{\dL^2}{2} + \ldots \\
{\rm e}^{s\mu/\sigma^2} &= 1 + s\mu_1\dL \notag\\
&\ph{1+}
+ \left(s^2\mu_1^2 + s\left(\mu_2-2\mu_1\Sigma_1\right)\right)\frac{\dL^2}{2} + \ldots \\
{\rm e}^{-\frac{s^2}{2}\left(\frac{1}{\sigma^2}-1\right)} &= 1 + \frac{s^2\Sigma_1}{2}\dL  \notag\\
&\ph{1+}
+\left(\frac{s^2}{2}\left(\Sigma_2-2\Sigma_1^2\right)+\frac{s^4\Sigma_1^2}{4}\right)\frac{\dL^2}{2} + \ldots
\end{align}
Plugging these into \eqn{eq:gaussratio} and recognising the Hermite polynomials \eqref{eq:Hermite-1to4} gives us
\begin{align}
\frac{p(s|m,\dL)}{p(s|m,\dL=0)} - 1 &= \dL\Big(\mu_1H_1(s) + \frac{\Sigma_1}{2}H_2(s)\Big) \notag\\
&\ph{1+}
+ \frac{\dL^2}{2}\bigg[\mu_2H_1(s)+\left(\mu_1^2+\frac{\Sigma_2}{2}\right)H_2(s) \notag\\
&\ph{1+\dL^2}
+\mu_1\Sigma_1H_3(s) + \frac{\Sigma_1^2}{4}H_4(s)\bigg]\,,
\end{align}
using which we can write
\begin{align}
\Cal{P}_1^{\rm L}(m,s) &= \mu_1(m)H_1(s) + \frac{\Sigma_1(m)}{2}H_2(s)\\
\Cal{P}_2^{\rm L}(m,s) &= \mu_2(m)H_1(s)+\left(\mu_1(m)^2+\frac{\Sigma_2(m)}{2}\right)H_2(s) \notag\\
&\ph{1+}
+\mu_1(m)\Sigma_1(m)H_3(s) + \frac{\Sigma_1(m)^2}{4}H_4(s)\,,
\end{align}
which in turn, upon using \eqn{eq:bnLag(m,s)}, leads to \eqns{eq:b1Lag(m,s)} and~\eqref{eq:b2Lag(m,s)}.

The reason why $b_n^{\rm L}$, in general, involves Hermite polynomials up to order $H_{2n}(s)$ can also be understood as follows. The coefficient $b_n^{\rm L}$ follows from $n$ derivatives with respect to \dL\ of the quantity ${\rm e}^{-y(\dL)^2/2}/\sigma(\dL)$ where $y(\dL)=(s-\mu(\dL))/\sigma(\dL)$. These derivatives lead to terms involving, among others, $H_n(s)(\p y/\p\dL)^n\mid_{\dL=0}$. This in turn will involve $s^nH_n(s)$ and terms with lower powers of $s$ multiplying $H_n(s)$. A straightforward repeated application of the recurrence $H_{n+1}(s) = sH_n(s)-nH_{n-1}(s)$ to the term $s^nH_n(s)$ then explains the appearance of $H_{2n}(s)$.

\label{lastpage}

\end{document}